\documentclass[a4paper,11pt]{article}
\usepackage{pos}
\usepackage{slashed}
\usepackage{tikz}
\usetikzlibrary{trees}
\usetikzlibrary{decorations.pathmorphing}
\usetikzlibrary{decorations.markings}
\usepackage{tikz,ifthen}
\tikzset{beamerprimary/.style={structure.fg, thick}}
\tikzset{beamersecondary/.style={structure.bg, thick}}
\tikzset{boson/.style={draw=structure.fg,decorate, decoration={snake}},
    gauge/.style={decorate, decoration={snake} },
    fermion/.style={postaction={decorate},
        decoration={markings,mark=at position .55 with {\arrow{>}}}},
    fermionloop/.style={postaction={decorate},
        decoration={markings,mark=at position .25 with {\arrow{<}}}}, 
    gluon/.style={decorate, 
        decoration={coil,amplitude=4pt, segment length=5pt}},
    scalar/.style={dashed},
    scalarloop/.style={dashed={decorate},
        decoration={markings,mark=at position .25 with {\arrow{<}}}},
    resonance/.style={double,double distance=1.5pt}	
}
\usetikzlibrary{arrows,shapes,positioning}
\usetikzlibrary{decorations.markings}
\tikzstyle arrowstyle=[scale=1]
\tikzstyle directed=[postaction={decorate,decoration={markings,
    mark=at position .65 with {\arrow[arrowstyle]{stealth}}}}]
\tikzstyle reverse directed=[postaction={decorate,decoration={markings,
    mark=at position .65 with {\arrowreversed[arrowstyle]{stealth};}}}]

\title{Isospin-breaking corrections to \texorpdfstring{$\tau^- \to \pi^- \pi^0 \nu_\tau$}{Lg} decays and the muon \texorpdfstring{$g-2$}{Lg}}

\author*[a]{Alejandro Miranda}

\affiliation[a]{Departamento de F\'isica, Centro de Investigaci\'on y de Estudios Avanzados del IPN,\\
  Apdo. Postal 14-740,07000 Ciudad de M\'exico, M\'exico}

\emailAdd{jmiranda@fis.cinvestav.mx}

\abstract{We review the isospin-breaking and electromagnetic corrections to the $\tau^- \to \pi^- \pi^0 \nu_\tau$ decays, which are used as an input to the two-pion contributions to the hadronic vacuum polarization (at LO) of the anomalous magnetic moment ($a_\mu$). We extend previous analyses by Cirigliano et al. working with ChPT with resonances. As an outcome, we improve the agreement between this determination and the other based on $e^+e^-$ data. The new results are in better agreement with an old estimation that uses Vector Dominance Model (VMD), and the discrepancy between the SM prediction and the combined results from BNL and FNAL is reduced to $2.1\,\sigma$ at $\mathcal{O}(p^4)$ and $2.3\sigma$ at $\mathcal{O}(p^6)$.}

\FullConference{%
  *** 10th International Workshop on Charm Physics (CHARM2020), ***\\
  *** 31 May - 4 June, 2021 ***\\
  *** Mexico City, Mexico - Online ***
}

\DeclareUnicodeCharacter{2212}{\textendash}

\begin{document}
\maketitle

\section{Introduction}
The Standard Model of particle physics \cite{Glashow:1959wxa,Salam:1959zz,Weinberg:1967tq} is one of the most successful theories that describes strong, weak and electromagnetic interactions. One of the most precisely measured quantities in particle physics is the muon anomalous magnetic moment ($a_\mu$). A long-standing discrepancy between theory and experiment about $3-4$ standard deviations has been observed.

The agreement between the latest measurement performed at Fermilab National Accelerator Laboratory (FNAL) Muon $g-2$ Experiment \cite{Muong-2:2021ojo} with the previous one at Brookhaven \cite{Bennett:2006fi}, allows to combine them and get
\begin{equation*}
    a_\mu^{\mathrm{Exp}}=116\,592\,061(41)\times 10^{-11} \quad (0.35\text{ ppm}).
\end{equation*}
The latest SM prediction \cite{Aoyama:2020ynm} is 
\begin{equation*}
    a_\mu^{\mathrm{SM}}=116\,591\,810(43)\times 10^{-11} \quad (0.37\text{ ppm}).
\end{equation*}
Therefore, the difference, $\Delta a_\mu=a_\mu^{\text{Exp}}-a_\mu^{\text{SM}}=(251\pm59)\times 10^{-11}$, increases the significance at $4.2\,\sigma$. This deviation from the Standard Model could be a sign of New Physics around the corner.

The uncertainty of the SM prediction is dominated by the hadronic contributions. Specifically, the HVP, LO contribution is dominated by the $\pi\pi$ cut (about $\sim 70\%$ of the overall value). At present, the most competitive estimation is obtained via dispersion relations together with $e^+e^-\to \text{hadrons}$ data. Alternatively, the CVC relation between electromagnetic and weak form factors in the isospin-limit allows using tau-data for this purpose. When tau data is employed for $a_\mu^{HVP, LO}$, the discrepancy between the SM prediction and the experimental measurement amounts to $2.4\,\sigma$ \cite{Davier:2013sfa}. The difference with respect to the $e^+e^-$-based evaluation could be owing to NP effects, hinting at a lepton universality violation in the corresponding non-standard vector and/or tensor couplings at low-energies \cite{Miranda:2018cpf, Cirigliano:2018dyk, Gonzalez-Solis:2019owk}. Nowadays, $e^+e^-$-based estimation has superseded the tau-based one due to the very high precision achieved in the $\sigma(e^+e^-\to \text{ hadrons})$ measurements.

Additionally to the data-driven approach, lattice QCD determinations of $a_\mu^{\text{HVP, LO}}$ have achieved a significant improvement. Although the lattice estimations are not yet competitive with the $e^+e^-$ evaluations \cite{Aoyama:2020ynm}, a very accurate computation made by the BMW coll. \cite{Borsanyi:2020mff} alleviates the tension concerning the SM prediction at one sigma level.

The paper is organized as follows. In Sec. \ref{tautopipi} we review the main features of the $\tau^-\to\pi^-\pi^0\nu_\tau\gamma$ decays as well as the theoretical framework. We show some decay observables in Sec. \ref{sec:decayobs}. Then, in Sec. \ref{sec:HVP} we evaluate $a_\mu^{\text{HVP, LO}\vert_{\pi\pi}}$ using tau data. Finally, our conclusions are presented in Sec. \ref{sec:conc}.

\section{\texorpdfstring{$\tau^-\to\pi^-\pi^0\gamma\nu_\tau$}{Lg} decays}\label{tautopipi}
\subsection{Amplitude}
For the radiative decay $\tau^-\left(P\right)\to \pi^-\left(p_-\right)\pi^0\left(p_0\right)\nu_\tau\left(q\right)\gamma\left(k\right)$, we can split the contribution due to the bremsstrahlung off the initial tau lepton from the one coming from the hadronic part.

The most general structure for these decays can be written as \cite{Bijnens:1992en, Cirigliano:2002pv}
\begin{equation}\begin{split}
T=e\, G_F V_{ud}^*\epsilon^{\mu}(k)^*&\left\lbrace F_\nu\, \bar{u}\left(q\right)\gamma^\nu\left(1-\gamma_5\right)\left(m_\tau+\slashed P-\slashed k\right)\gamma_\mu\, u\left(P\right)\right.\\
&\left. +\left(V_{\mu\nu}-A_{\mu\nu}\right)\bar{u}\left(q\right)\gamma^\nu\left(1-\gamma_5\right)u\left(P\right)\right\rbrace,
\end{split}\end{equation}
where $F_\nu\equiv \left(p_0-p_-\right)_\nu f_+\left(s\right)/2P\cdot k$, with the charged pion vector form factor $f_+(s)$ defined through $\left\langle\pi^0\pi^-|\bar{d}\gamma^\mu u|0\right\rangle=\sqrt{2}f_+(s)(p_{-}-p_0)^\mu$ and $s=(p_{-}+p_0)^2$. Gauge invariance ($\epsilon_\mu\to \epsilon_\mu+k_\mu$) implies the Ward identities 

\begin{equation}\label{WardIds}
k_\mu V^{\mu\nu}=\left(p_{-}-p_0\right)^\nu f_+\left(s\right),\quad k_\mu A^{\mu\nu}=0.
\end{equation}

Imposing eq.~(\ref{WardIds}) and Lorentz invariance, we have the following expression for the vector structure-dependent tensor
\begin{equation}\label{FFW:V}\small\begin{split}
V^{\mu\nu}&=f_+\left[\left(P-q\right)^2\right] \frac{p_-^\mu \left(p_{-}+k -p_0\right)^\nu}{p_-\cdot k}-f_+\left[\left(P-q\right)^2\right]g^{\mu\nu}\\
&+\frac{f_+\left[\left(P-q\right)^2\right]-f_+\left(s\right)}{\left(p_0+p_-\right)\cdot k}\left(p_0+p_-\right)^\mu\left(p_0-p_-\right)^\nu\\
&+v_1\left(g^{\mu\nu}\,p_-\cdot k-p_-^\mu k^\nu\right)+v_2\left(g^{\mu\nu}\,p_0\cdot k-p_0^\mu k^\nu\right)\\
&+v_3\left(p_0\cdot k\, p_-^\mu-p_-\cdot k\,p_0^\mu\right)p_-^\nu+v_4\left(p_0\cdot k\, p_-^\mu-p_-\cdot k\,p_0^\mu\right)\left(p_0+p_-+k\right)^\nu,
\end{split}\end{equation}
and for the axial one
\begin{equation}\label{FFW:A}\begin{split}
A^{\mu\nu} &=ia_1\,\epsilon^{\mu\nu\rho\sigma}\,\left(p_0-p_-\right)_{\rho}k_{\sigma}+ia_2\,W^\nu\,\epsilon^{\mu\lambda\rho\sigma}k_{\lambda}\,p_{-\rho}\,p_{0\sigma}\\
&+ia_3\,\epsilon^{\mu\nu\rho\sigma}k_{\rho}\,W_{\sigma}+ia_4\,\left(p_0+k\right)^\nu\,\epsilon^{\mu\lambda\rho\sigma}\,k_\lambda\,p_{-\rho}\,p_{0\sigma},\\
\end{split}\end{equation}
where $W\equiv P-q=p_-+p_0+k$. The structure-dependent contributions to these tensor structures are contained in the four vector ($v_i$) and the four axial-vector ($a_i$) form factors. For the axial structure, the Schouten's identity has been used.

Using $\left(P-q\right)^2=s+2\left(p_0+p_-\right)\cdot k$, it is easy to show that Low's theorem \cite{Low:1958sn} is manifestly satisfied  
\begin{equation}\label{eq:Lowapp}\begin{split}
V^{\mu\nu}&=f_+\left(s\right) \frac{p_-^\mu }{p_-\cdot k}\left(p_{-}-p_0\right)^\nu +f_+\left(s\right)\left(\frac{p_-^\mu k^\nu}{p_-\cdot k}-g^{\mu\nu} \right)\\
&+2\frac{df_+\left(s\right)}{d\,s}\left(\frac{p_0\cdot k}{p_-\cdot k}p_-^\mu-p_0^\mu\right)\left(p_{-}-p_0\right)^\nu+\mathcal{O}\left(k\right).
\end{split}\end{equation}

\subsection{Vector Form Factors}\label{ss:VFF}
In $R\chi T$ \cite{Ecker:1988te, Ecker:1989yg, Cirigliano:2006hb, Kampf:2011ty}, the diagrams that contribute to the vector form factors of the $\tau^-\to\pi^-\pi^0\gamma\nu_\tau$ decays are shown in Figs. \ref{VF:fig1}, \ref{VF:fig2} and \ref{VF:fig3}.

\begin{figure}[!ht]
\begin{center}
\begin{tikzpicture}
	\draw[resonance] (-1.5,0)--(0,0);
	\node at (-0.6,-0.2) {\tiny $\rho^-$};
	\draw[fill,white] (-1.5,0) circle [radius=0.12];
	\node at (-1.5,0) {\small$\otimes$};
	\draw[scalar] (0,0)--(0.75,0.75) node[right] {\tiny$\pi^-$};
	\draw[gauge] (0,0)--(0.75,0) node[right] {\tiny $\gamma$};
	\draw[scalar] (0,0)--(0.75,-0.75) node[right] {\tiny $\pi^0$};
	\draw[fill] (0,0) circle [radius=0.04];
\end{tikzpicture}
\begin{tikzpicture}
	\draw[resonance] (2,0)--(3.5,0);
	\node at (2.9,-0.2) {\tiny$\rho^-$};
	\draw[gauge] (2,0)--(2,0.75) node[above] {\tiny$\gamma$};
	\draw[fill,white] (2,0) circle [radius=0.12];
	\node at (2,0) {\small$\otimes$};
	\draw[scalar] (3.5,0)--(4.25,0.75) node[right] {\tiny$\pi^-$};
	\draw[scalar] (3.5,0)--(4.25,-0.75) node[right] {\tiny$\pi^0$};
	\draw[fill] (3.5,0) circle [radius=0.04];
\end{tikzpicture}
\begin{tikzpicture}
	\draw[resonance] (-1.5,0)--(0,0);
	\node at (-0.6,-0.2) {\tiny $\rho^-$};
	\draw[fill,white] (-1.5,0) circle [radius=0.12];
	\node at (-1.5,0) {\small$\otimes$};
	\draw[scalar] (0,0)--(0.75,0.75) node[right] {\tiny$\pi^-$};
	\draw[gauge] (0.375,0.375)--(0.8,-0.05) node[right] {\tiny $\gamma$};
	\draw[scalar] (0,0)--(0.75,-0.75) node[right] {\tiny $\pi^0$};
	\draw[fill] (0,0) circle [radius=0.04];
\end{tikzpicture}
\begin{tikzpicture}

	\draw[resonance] (5.5,0)--(6.7,0);
	\node at (6.2,-0.2) {\tiny$\rho^0$};
	\draw[scalar] (5.5,0)--(6.25,0.75) node[right] {\tiny$\pi^-$};
	\draw[scalar] (5.5,0)--(6.25,-0.75) node[right] {\tiny $\pi^0$};
	\draw[fill,white] (5.5,0) circle [radius=0.12];
	\node at (5.5,0) {\small $\otimes$};
	\draw[gauge] (6.7,0)--(7.5,0) node[right] {\tiny$\gamma$};
	\draw[fill] (6.7,0) circle [radius=0.04];
\end{tikzpicture}
\begin{tikzpicture}
	\draw[resonance] (2,-2)--(3.5,-2);
	\node at (2.8,-2.2) {\tiny $\omega$};
	\draw[scalar] (2,-2)--(2,-1.25) node[above] {\tiny$\pi^-$};
	\draw[fill,white] (2,-2) circle [radius=0.12];
	\node at (2,-2) {\small$\otimes$};
	\draw[scalar] (3.5,-2)--(4.25,-1.25) node[right] {\tiny$\pi^0$};
	\draw[gauge] (3.5,-2)--(4.25,-2.75) node[right] {\tiny$\gamma$};
	\draw[fill] (3.5,-2) circle [radius=0.04];
\end{tikzpicture}
\begin{tikzpicture}
	\draw[resonance] (-1.5,0)--(0,0);
	\node at (-0.6,-0.2) {\tiny $a_1^-$};
	\draw[scalar] (-1.5,0)--(-1.5,0.75) node[above] {\tiny $\pi^0$};
	\draw[fill,white] (-1.5,0) circle [radius=0.12];
	\node at (-1.5,0) {\small$\otimes$};
	\draw[scalar] (0,0)--(0.75,0.75) node[right] {\tiny$\pi^-$};
	\draw[gauge] (0,0)--(0.75,-0.75) node[right] {\tiny $\gamma$};
	\draw[fill] (0,0) circle [radius=0.04];
\end{tikzpicture}
\begin{tikzpicture}
	\draw[scalar] (8.5,-4)--(9.25,-4);
	\node at (8.99,-4.2) {\tiny $\pi^-$};
	\draw[resonance] (9.25,-4)--(10,-4);
	\node at (9.65,-4.2) {\tiny $\rho^0$};
	\draw[scalar] (8.5,-4)--(8.5,-3.25) node[above] {\tiny$\pi^0$};
	\draw[fill,white] (8.5,-4) circle [radius=0.12];
	\node at (8.5,-4) {\small $\otimes$};
	\draw[scalar] (9.25,-4)--(9.25,-3.25) node[above] {\tiny$\pi^-$};
	\draw[fill] (9.25,-4) circle [radius=0.04];
	\draw[gauge] (10,-4)--(10.75,-4) node[right] {\tiny$\gamma$};
	\draw[fill] (10,-4) circle [radius=0.04];
\end{tikzpicture}
\end{center}
\caption{One-resonance exchange contributions from the $R\chi T$ to the vector form factors of the $\tau^-\to\pi^-\pi^0\gamma\nu_\tau$ decays.}\label{VF:fig1}
\end{figure}
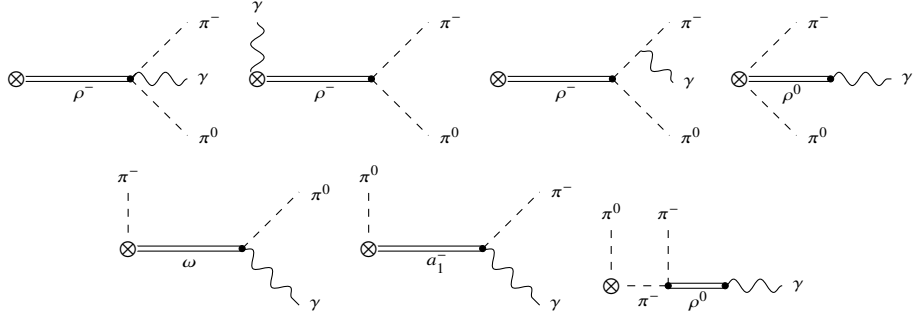

\begin{figure}[ht]
\begin{center}
\begin{tikzpicture}
	\draw[resonance] (-1.5,0)--(0,0);
	\node at (-1.01,-0.2) {\tiny $\rho^-$};
	\node at (-0.35,-0.2) {\tiny $\rho^-$};
	\draw[fill,white] (-1.5,0) circle [radius=0.12];
	\node at (-1.5,0) {\small$\otimes$};
	\draw[gauge] (-0.75,0)--(-0.75,0.75) node[above] {\tiny $\gamma$};
	\draw[fill] (-0.75,0) circle [radius=0.04];
	\draw[scalar] (0,0)--(0.75,0.75) node[right] {\tiny$\pi^-$};
	\draw[scalar] (0,0)--(0.75,-0.75) node[right] {\tiny $\pi^0$};
	\draw[fill] (0,0) circle [radius=0.04];
\end{tikzpicture}
\begin{tikzpicture}
	\draw[resonance] (2,0)--(3.5,0);
	\node at (2.49,-0.2) {\tiny$\rho^-$};
	\node at (3.25,-0.2) {\tiny$\rho^0$};
	\draw[fill,white] (2,0) circle [radius=0.12];
	\node at (2,0) {\small$\otimes$};
	\draw[scalar] (2.75,0)--(3.5,0.75) node[right] {\tiny$\pi^-$};
	\draw[scalar] (2.75,0)--(3.5,-0.75) node[right] {\tiny$\pi^0$};
	\draw[fill] (2.75,0) circle [radius=0.04];	
	\draw[gauge] (3.5,0)--(4.25,0) node[right] {\tiny$\gamma$};
	\draw[fill] (3.5,0) circle [radius=0.04];	
\end{tikzpicture}
\begin{tikzpicture}
	\draw[resonance] (5.5,0)--(7,0);
	\node at (5.99,-0.2) {\tiny$\rho^-$};
	\node at (6.65,-0.2) {\tiny$\omega$};
	\draw[fill,white] (5.5,0) circle [radius=0.12];
	\node at (5.5,0) {\small$\otimes$};
	\draw[scalar] (6.25,0)--(6.25,0.75) node[above] {\tiny$\pi^-$};
	\draw[fill] (6.25,0) circle [radius=0.04];
	\draw[scalar] (7,0)--(7.75,0.75) node[right] {\tiny$\pi^0$};
	\draw[gauge] (7,0)--(7.75,-0.75) node[right] {\tiny$\gamma$};
	\draw[fill] (7,0) circle [radius=0.04];
\end{tikzpicture}
\begin{tikzpicture}
	\draw[resonance] (2,-2)--(3.5,-2);
	\node at (3.15,-2.2) {\tiny$\rho^0$};
	\node at (2.4,-2.2) {\tiny$\omega$};
	\draw[scalar] (2,-2)--(2,-1.25) node[above] {\tiny$\pi^-$};
	\draw[fill,white] (2,-2) circle [radius=0.12];
	\node at (2,-2) {\small$\otimes$};
	\draw[scalar] (2.75,-2)--(2.75,-1.25) node[above] {\tiny$\pi^0$};
	\draw[fill] (2.75,-2) circle [radius=0.04];
	\draw[gauge] (3.5,-2)--(4.25,-2) node[right] {\tiny$\gamma$};
	\draw[fill] (3.5,-2) circle [radius=0.04];
\end{tikzpicture}
\begin{tikzpicture}
	\draw[resonance] (5.5,-2)--(6.8,-1.625);
	\node at (6.15,-1.62) {\tiny$\rho^0$};
	\draw[gauge] (6.8,-1.625)--(7.7,-1.25) node[right] {\tiny$\gamma$};
	\draw[fill] (6.8,-1.625) circle [radius=0.04];
	\draw[resonance] (5.5,-2)--(6.9,-2.375);
	\node at (6.15,-2.37) {\tiny$\rho^-$};
	\draw[fill,white] (5.5,-2) circle [radius=0.12];
	\node at (5.5,-2) {\small$\otimes$};
	\draw[scalar] (6.9,-2.375)--(7.75,-2) node[right] {\tiny$\pi^-$};
	\draw[scalar] (6.9,-2.375)--(7.75,-2.75) node[right] {\tiny$\pi^0$};
	\draw[fill] (6.9,-2.375) circle [radius=0.04];
\end{tikzpicture}
\begin{tikzpicture}
	\draw[resonance] (-1.5,0)--(0,0);
	\node at (-1.01,-0.2) {\tiny $\rho^-$};
	\node at (-0.35,-0.2) {\tiny $a_1^-$};
	\draw[fill,white] (-1.5,0) circle [radius=0.12];
	\node at (-1.5,0) {\small$\otimes$};
	\draw[scalar] (-0.75,0)--(-0.75,0.75) node[above] {\tiny $\pi^0$};
	\draw[fill] (-0.75,0) circle [radius=0.04];
	\draw[scalar] (0,0)--(0.75,0.75) node[right] {\tiny$\pi^-$};
	\draw[gauge] (0,0)--(0.75,-0.75) node[right] {\tiny $\gamma$};
	\draw[fill] (0,0) circle [radius=0.04];
\end{tikzpicture}
\begin{tikzpicture}
	\draw[resonance] (1.75,-2)--(3.25,-2);
	\draw[scalar] (1.75,-2)--(1.75,-1.25) node[above] {\tiny$\pi^0$};
	\node at (2.24,-2.2) {\tiny $a_1^-$};
	\node at (2.9,-2.2) {\tiny $\rho^0$};
	\draw[fill,white] (1.75,-2) circle [radius=0.12];
	\node at (1.75,-2) {\small$\otimes$};
	\draw[scalar] (2.5,-2)--(2.5,-1.25) node[above] {\tiny $\pi^-$};
	\draw[fill] (2.5,-2) circle [radius=0.04];
	\draw[gauge] (3.25,-2)--(4,-2) node[right] {\tiny $\gamma$};
	\draw[fill] (3.25,-2) circle [radius=0.04];
\end{tikzpicture}
\begin{tikzpicture}
	\draw[resonance] (7.75,-4)--(8.5,-4);
	\draw[fill,white] (7.75,-4) circle [radius=0.12];
	\node at (7.75,-4) {\small$\otimes$};
	\node at (8.125,-4.2) {\tiny$\rho^-$};
	\draw[scalar] (8.5,-4)--(8.5,-3.25) node[above] {\tiny$\pi^0$};
	\draw[scalar] (8.5,-4)--(9.25,-4);
	\draw[resonance] (9.25,-4)--(10,-4);
	\draw[fill] (8.5,-4) circle [radius=0.04];
	\node at (8.875,-4.2) {\tiny$\pi^-$};
	\node at (9.625,-4.2) {\tiny$\rho^0$};
	\draw[scalar] (9.25,-4)--(9.25,-3.25) node[above] {\tiny$\pi^-$};
	\draw[fill] (9.25,-4) circle [radius=0.04];
	\draw[gauge] (10,-4)--(10.75,-4) node[right] {\tiny$\gamma$};
	\draw[fill] (10,-4) circle [radius=0.04];
\end{tikzpicture}
\end{center}
\caption{Two-resonance exchange contributions from the $R\chi T$ to the vector form factors of the $\tau^-\to\pi^-\pi^0\gamma\nu_\tau$ decays.}\label{VF:fig2}
\end{figure}
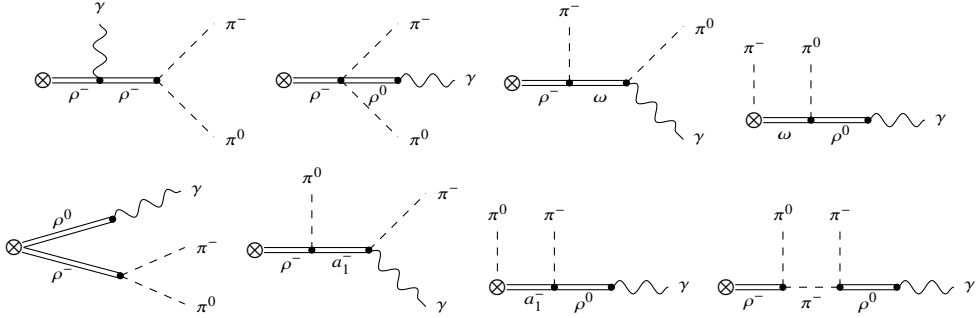

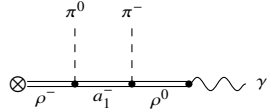
\begin{figure}[ht]
\begin{center}
\begin{tikzpicture}
	\draw[resonance] (7.75,-4)--(10,-4);
	\draw[fill,white] (7.75,-4) circle [radius=0.12];
	\node at (7.75,-4) {\small$\otimes$};
	\node at (8.125,-4.2) {\tiny$\rho^-$};
	\draw[scalar] (8.5,-4)--(8.5,-3.25) node[above] {\tiny$\pi^0$};
	\draw[fill] (8.5,-4) circle [radius=0.04];
	\node at (8.875,-4.2) {\tiny$a_1^-$};
	\node at (9.625,-4.2) {\tiny$\rho^0$};
	\draw[scalar] (9.25,-4)--(9.25,-3.25) node[above] {\tiny$\pi^-$};
	\draw[fill] (9.25,-4) circle [radius=0.04];
	\draw[gauge] (10,-4)--(10.75,-4) node[right] {\tiny$\gamma$};
	\draw[fill] (10,-4) circle [radius=0.04];

\end{tikzpicture}
\end{center}
\caption{Three-resonance exchange contributions from the $R\chi T$ to the vector form factors of the $\tau^-\to\pi^-\pi^0\gamma\nu_\tau$ decays.}\label{VF:fig3}
\end{figure}

For the vector form factors, we have
\begin{subequations}\label{eqsvs}\small\begin{align}
v_1&=v_1^0+v_1^R+v_1^{RR}+v_1^{RRR}+v_{GI1}^{R+RR},\\
v_2&=v_2^0+v_2^R+v_2^{RR}+v_2^{RRR}+v_{GI2}^{R+RR},\\
v_3&=v_3^0+v_3^R+v_3^{RR}+v_3^{RRR}+v_{GI3}^{R+RR},\\
v_4&=v_4^0+v_4^R+v_4^{RR}+v_4^{RRR}+v_{GI4}^{R+RR},
\end{align}\end{subequations}
where $v_i^0$ is the $\mathcal{O}(p^4)$ contribution in Ref. \cite{Cirigliano:2002pv},
and $v_i^{R}$, $v_i^{RR}$, $v_i^{RRR}$ and $v_{GIi}^{R+RR}$, which are the subleading contributions up to $\mathcal{O}(p^6)$, can be found in App. C. in Ref. \cite{Miranda:2020wdg}.

\subsection{Axial-Vector Form Factors}\label{subsec:AVFFs}
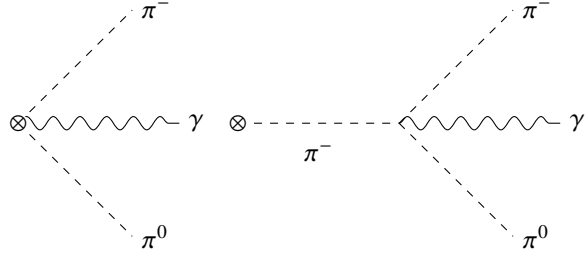
\begin{figure}[ht]
\begin{center}{\small
\begin{tikzpicture}
	\draw[scalar] (0,0) -- (1.5,1.5) node [right]{$\pi^-$} ;
	\draw[gauge] (0,0) -- (2.12,0) node[right]{$\gamma$};
	\draw[scalar] (0,0) -- (1.5,-1.5) node [right] {$\pi^0$};
	\draw[fill,white] (0,0) circle [radius=0.12];
	\node at (0,0) {\small$\otimes$};
\end{tikzpicture}
\begin{tikzpicture}
	\draw[scalar] (5.62,0)--(7.74,0);
	\node at (6.68,-0.4) {$\pi^-$};
	\draw[scalar] (7.74,0)--(9.24,1.5) node [right]{$\pi^-$} ;
	\draw[gauge] (7.74,0)--(9.86,0) node[right]{$\gamma$};
	\draw[scalar] (7.74,0)--(9.24,-1.5) node [right] {$\pi^0$};
	\draw[fill,white] (5.62,0) circle [radius=0.12];
	\node at (5.62,0) {\small$\otimes$};
\end{tikzpicture}}
\end{center}
\caption{Anomalous diagrams contributing to the axial tensor amplitude $A^{\mu\nu}$ at $\mathcal{O}\left(p^4\right)$.}\label{Ap:fig3}
\end{figure}

The axial form factors at chiral $\mathcal{O}\left(p^4\right)$ \cite{Cirigliano:2002pv} get contributions from the Wess-Zumino-Witten functional \cite{Wess:1971yu,Witten:1983tw}:
\begin{equation}
a_1^0\equiv\frac{1}{8\pi^2 F^2},\qquad a_2^0\equiv\frac{-1}{4\pi^2 F^2\left[\left(P-q\right)^2-m_\pi^2\right]}.
\end{equation}
The diagrams contributing to these two expressions are shown in fig. \ref{Ap:fig3}.

\begin{figure}[ht]
\begin{center}
\begin{tikzpicture}
	\draw[resonance] (2,0)--(3.5,0);
	\node at (2.9,-0.2) {\tiny$\rho^-$};
	\draw[gauge] (2,0)--(2,0.75) node[above] {\tiny$\gamma$};
	\draw[fill,white] (2,0) circle [radius=0.12];
	\node at (2,0) {\small$\otimes$};
	\draw[scalar] (3.5,0)--(4.25,0.75) node[right] {\tiny$\pi^-$};
	\draw[scalar] (3.5,0)--(4.25,-0.75) node[right] {\tiny$\pi^0$};
	\draw[fill] (3.5,0) circle [radius=0.04];
\end{tikzpicture}
\begin{tikzpicture}
	\draw[resonance] (8.5,0)--(10,0);
	\node at (9.4,-0.2) {\tiny$\rho^-$};
	\draw[scalar] (8.5,0)--(8.5,0.75) node[above] {\tiny$\pi^0$};
	\draw[fill,white] (8.5,0) circle [radius=0.12];
	\node at (8.5,0) {\small$\otimes$};
	\draw[scalar] (10,0)--(10.75,0.75) node[right] {\tiny$\pi^-$};
	\draw[gauge] (10,0)--(10.75,-0.75) node[right] {\tiny$\gamma$};
	\draw[fill] (10,0) circle [radius=0.04];
\end{tikzpicture}
\begin{tikzpicture}
	\draw[resonance] (-1.5,-2)--(0,-2);
	\node at (-0.6,-2.2) {\tiny $\rho^0$};
	\draw[scalar] (-1.5,-2)--(-1.5,-1.25) node[above] {\tiny$\pi^-$};
	\draw[fill,white] (-1.5,-2) circle [radius=0.12];
	\node at (-1.5,-2) {\small$\otimes$};
	\draw[scalar] (0,-2)--(0.75,-1.25) node[right] {\tiny$\pi^0$};
	\draw[gauge] (0,-2)--(0.75,-2.75) node[right] {\tiny$\gamma$};
	\draw[fill] (0,-2) circle [radius=0.04];
\end{tikzpicture}
\begin{tikzpicture}
	\draw[resonance] (5.5,-2)--(6.7,-2);
	\node at (6.2,-2.2) {\tiny$\omega$};
	\draw[scalar] (5.5,-2)--(6.25,-1.25) node[right] {\tiny$\pi^-$};
	\draw[scalar] (5.5,-2)--(6.25,-2.75) node[right] {\tiny $\pi^0$};
	\draw[fill,white] (5.5,-2) circle [radius=0.12];
	\node at (5.5,-2) {\small $\otimes$};
	\draw[gauge] (6.7,-2)--(7.5,-2) node[right] {\tiny$\gamma$};
	\draw[fill] (6.7,-2) circle [radius=0.04];
\end{tikzpicture}
\begin{tikzpicture}
	\draw[resonance] (9.25,-2)--(10,-2);
	\node at (9.65,-2.2) {\tiny $\rho^0$};
	\draw[scalar] (8.5,-2)--(9.25,-2);
	\node at (8.99,-2.2) {\tiny $\pi^-$};
	\draw[fill] (9.25,-2) circle [radius=0.04];
	\draw[fill,white] (8.5,-2) circle [radius=0.12];
	\node at (8.5,-2) {\small $\otimes$};
	\draw[scalar] (9.25,-2)--(9.25,-1.25) node[above] {\tiny $\pi^-$};
	\draw[scalar] (10,-2)--(10.75,-1.25) node[right] {\tiny $\pi^0$};
	\draw[gauge] (10,-2)--(10.75,-2.75) node[right] {\tiny $\gamma$};
	\draw[fill] (10,-2) circle [radius=0.04];
\end{tikzpicture}
\begin{tikzpicture}
	\draw[scalar] (-1.5,-4)--(-0.75,-4);
	\node at (-1.01,-4.2) {\tiny $\pi^-$};
	\draw[resonance] (-0.75,-4)--(0,-4);
	\node at (-0.35,-4.2) {\tiny $\rho^-$};
	\draw[fill,white] (-1.5,-4) circle [radius=0.12];
	\node at (-1.5,-4) {\small $\otimes$};
	\draw[scalar] (-0.75,-4)--(-0.75,-3.25) node[above] {\tiny$\pi^0$};
	\draw[fill] (-0.75,-4) circle [radius=0.04];
	\draw[scalar] (0,-4)--(0.75,-3.25) node[right] {\tiny$\pi^-$};
	\draw[gauge] (0,-4)--(0.75,-4.75) node[right] {\tiny$\gamma$};
	\draw[fill] (0,-4) circle [radius=0.04];
\end{tikzpicture}
\begin{tikzpicture}
	\draw[scalar] (2,-4)--(2.75,-4);
	\node at (2.49,-4.2) {\tiny $\pi^-$};
	\draw[resonance] (2.75,-4)--(3.5,-4);
	\node at (3.15,-4.2) {\tiny $\rho^-$};
	\draw[fill,white] (2,-4) circle [radius=0.12];
	\node at (2,-4) {\small $\otimes$};
	\draw[gauge] (2.75,-4)--(2.75,-3.25) node[above] {\tiny$\gamma$};
	\draw[fill] (2.75,-4) circle [radius=0.04];
	\draw[scalar] (3.5,-4)--(4.25,-3.25) node[right] {\tiny$\pi^-$};
	\draw[scalar] (3.5,-4)--(4.25,-4.75) node[right] {\tiny$\pi^0$};
	\draw[fill] (3.5,-4) circle [radius=0.04];
\end{tikzpicture}
\begin{tikzpicture}
	\draw[scalar] (5.5,-4)--(6.25,-4);
	\node at (5.99,-4.2) {\tiny $\pi^-$};
	\draw[resonance] (6.25,-4)--(7,-4);
	\node at (6.69,-4.2) {\tiny $\omega$};
	\draw[fill,white] (5.5,-4) circle [radius=0.12];
	\node at (5.5,-4) {\small $\otimes$};
	\draw[scalar] (6.25,-4)--(7,-3.25) node[right] {\tiny$\pi^-$};
	\draw[scalar] (6.25,-4)--(7,-4.75) node[right] {\tiny$\pi^0$};
	\draw[fill] (6.25,-4) circle [radius=0.04];
	\draw[gauge] (7,-4)--(7.7,-4) node[right] {\tiny$\gamma$};
	\draw[fill] (7,-4) circle [radius=0.04];
\end{tikzpicture}
\begin{tikzpicture}
	\draw[resonance] (2,0)--(3.5,0);
	\node at (2.9,-0.2) {\tiny $a_1^-$};
	\draw[fill,white] (2,0) circle [radius=0.12];
	\node at (2,0) {\small$\otimes$};
	\draw[scalar] (3.5,0)--(4.25,0.75) node[right] {\tiny$\pi^-$};
	\draw[gauge] (3.5,0)--(4.25,0) node[right] {\tiny $\gamma$};
	\draw[scalar] (3.5,0)--(4.25,-0.75) node[right] {\tiny $\pi^0$};
	\draw[fill] (3.5,0) circle [radius=0.04];
\end{tikzpicture}
\end{center}
\caption{One-resonance exchange contributions from the $R\chi T$ to the axial-vector form factors of the $\tau^-\to\pi^-\pi^0\gamma\nu_\tau$ decays.}\label{AF:fig1}
\end{figure}
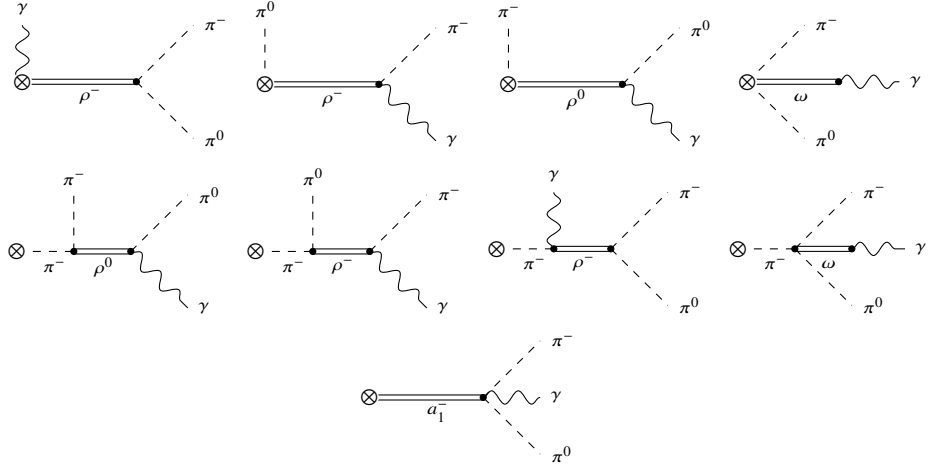

\begin{figure}[ht]
\begin{center}
\begin{tikzpicture}
	\draw[resonance] (9,0)--(10.5,0);
	\node at (9.4,-0.2) {\tiny$\rho^0$};
	\node at (10.15,-0.2) {\tiny$\omega$};
	\draw[scalar] (9,0)--(9,0.75) node[above] {\tiny$\pi^-$};
	\draw[fill,white] (9,0) circle [radius=0.12];
	\node at (9,0) {\small$\otimes$};
	\draw[scalar] (9.75,0)--(9.75,0.75) node[above] {\tiny$\pi^0$};
	\draw[fill] (9.75,0) circle [radius=0.04];
	\draw[gauge] (10.5,0)--(11.25,0) node[right] {\tiny$\gamma$};
	\draw[fill] (10.5,0) circle [radius=0.04];
\end{tikzpicture}
\begin{tikzpicture}
	\draw[resonance] (-1.5,-2)--(0,-2);
	\node at (-1.1,-2.2) {\tiny$\rho^-$};
	\node at (-0.35,-2.2) {\tiny$\omega$};
	\draw[scalar] (-1.5,-2)--(-1.5,-1.25) node[above] {\tiny$\pi^0$};
	\draw[fill,white] (-1.5,-2) circle [radius=0.12];
	\node at (-1.5,-2) {\small$\otimes$};
	\draw[scalar] (-0.75,-2)--(-0.75,-1.25) node[above] {\tiny$\pi^-$};
	\draw[fill] (-0.75,-2) circle [radius=0.04];
	\draw[gauge] (0,-2)--(0.75,-2) node[right] {\tiny$\gamma$};
	\draw[fill] (0,-2) circle [radius=0.04];
\end{tikzpicture}
\begin{tikzpicture}
	\draw[resonance] (9,-2)--(10.3,-1.625);
	\node at (9.65,-1.62) {\tiny$\omega$};
	\draw[gauge] (10.3,-1.625)--(11.2,-1.25) node[right] {\tiny$\gamma$};
	\draw[fill] (10.3,-1.625) circle [radius=0.04];
	\draw[resonance] (9,-2)--(10.4,-2.375);
	\node at (9.65,-2.37) {\tiny$\rho^-$};
	\draw[fill,white] (9,-2) circle [radius=0.12];
	\node at (9,-2) {\small$\otimes$};
	\draw[scalar] (10.4,-2.375)--(11.25,-2) node[right] {\tiny$\pi^-$};
	\draw[scalar] (10.4,-2.375)--(11.25,-2.75) node[right] {\tiny$\pi^0$};
	\draw[fill] (10.4,-2.375) circle [radius=0.04];
\end{tikzpicture}
\begin{tikzpicture}
	\draw[scalar] (-0.75,-4)--(0,-4);
	\draw[fill,white] (-0.75,-4) circle [radius=0.12];
	\node at (-0.75,-4) {\small$\otimes$};
	\node at (-0.375,-4.2) {\tiny$\pi^-$};
	\draw[scalar] (0,-4)--(0,-3.25) node[above] {\tiny$\pi^-$};
	\draw[resonance] (0,-4)--(1.5,-4);
	\draw[fill] (0,-4) circle [radius=0.04];
	\node at (0.375,-4.2) {\tiny$\rho^0$};
	\node at (1.125,-4.2) {\tiny$\omega$};
	\draw[scalar] (0.75,-4)--(0.75,-3.25) node[above] {\tiny$\pi^0$};
	\draw[fill] (0.75,-4) circle [radius=0.04];
	\draw[gauge] (1.5,-4)--(2.25,-4) node[right] {\tiny$\gamma$};
	\draw[fill] (1.5,-4) circle [radius=0.04];
\end{tikzpicture}
\begin{tikzpicture}
	\draw[scalar] (3.5,-4)--(4.25,-4);
	\draw[fill,white] (3.5,-4) circle [radius=0.12];
	\node at (3.5,-4) {\small$\otimes$};
	\node at (3.875,-4.2) {\tiny$\pi^-$};
	\draw[scalar] (4.25,-4)--(4.25,-3.25) node[above] {\tiny$\pi^0$};
	\draw[resonance] (4.25,-4)--(5.75,-4);
	\draw[fill] (4.25,-4) circle [radius=0.04];
	\node at (4.625,-4.2) {\tiny$\rho^-$};
	\node at (5.375,-4.2) {\tiny$\omega$};
	\draw[scalar] (5,-4)--(5,-3.25) node[above] {\tiny$\pi^-$};
	\draw[fill] (5,-4) circle [radius=0.04];
	\draw[gauge] (5.75,-4)--(6.5,-4) node[right] {\tiny$\gamma$};
	\draw[fill] (5.75,-4) circle [radius=0.04];
\end{tikzpicture}
\begin{tikzpicture}
	\draw[resonance] (1.75,0)--(3.25,0);
	\node at (2.24,-0.2) {\tiny $a_1^-$};
	\node at (2.9,-0.2) {\tiny $\rho^-$};
	\draw[fill,white] (1.75,0) circle [radius=0.12];
	\node at (1.75,0) {\small$\otimes$};
	\draw[scalar] (2.5,0)--(2.5,0.75) node[above] {\tiny $\pi^0$};
	\draw[fill] (2.5,0) circle [radius=0.04];
	\draw[scalar] (3.25,0)--(4,0.75) node[right] {\tiny$\pi^-$};
	\draw[gauge] (3.25,0)--(4,-0.75) node[right] {\tiny $\gamma$};
	\draw[fill] (3.25,0) circle [radius=0.04];
\end{tikzpicture}
\begin{tikzpicture}
	\draw[resonance] (5,0)--(6.5,0);
	\node at (5.49,-0.2) {\tiny $a_1^-$};
	\node at (6.15,-0.2) {\tiny $\rho^0$};
	\draw[fill,white] (5,0) circle [radius=0.12];
	\node at (5,0) {\small$\otimes$};
	\draw[scalar] (5.75,0)--(5.75,0.75) node[above] {\tiny $\pi^-$};
	\draw[fill] (5.75,0) circle [radius=0.04];
	\draw[scalar] (6.5,0)--(7.25,0.75) node[right] {\tiny$\pi^0$};
	\draw[gauge] (6.5,0)--(7.25,-0.75) node[right] {\tiny $\gamma$};
	\draw[fill] (6.5,0) circle [radius=0.04];
\end{tikzpicture}
\begin{tikzpicture}
	\draw[resonance] (8.25,0)--(9.75,0);
	\node at (8.74,-0.2) {\tiny $a_1^-$};
	\node at (9.4,-0.2) {\tiny $\rho^-$};
	\draw[fill,white] (8.25,0) circle [radius=0.12];
	\node at (8.25,0) {\small$\otimes$};
	\draw[gauge] (9,0)--(9,0.75) node[above] {\tiny $\gamma$};
	\draw[fill] (9,0) circle [radius=0.04];
	\draw[scalar] (9.75,0)--(10.5,0.75) node[right] {\tiny$\pi^-$};
	\draw[scalar] (9.75,0)--(10.5,-0.75) node[right] {\tiny $\pi^0$};
	\draw[fill] (9.75,0) circle [radius=0.04];
\end{tikzpicture}
\begin{tikzpicture}
	\draw[resonance] (5,-2)--(6.5,-2);
	\node at (5.49,-2.2) {\tiny $a_1^-$};
	\node at (6.2,-2.2) {\tiny $\omega$};
	\draw[fill,white] (5,-2) circle [radius=0.12];
	\node at (5,-2) {\small$\otimes$};
	\draw[scalar] (5.75,-2)--(6.5,-1.25) node[right] {\tiny$\pi^-$};
	\draw[scalar] (5.75,-2)--(6.5,-2.75) node[right] {\tiny $\pi^0$};
	\draw[fill] (5.75,-2) circle [radius=0.04];
	\draw[gauge] (6.5,-2)--(7.25,-2) node[right] {\tiny $\gamma$};
	\draw[fill] (6.5,-2) circle [radius=0.04];
\end{tikzpicture}

\end{center}
\caption{Two-resonance exchange contributions from the $R\chi T$ to the axial-vector form factors of the $\tau^-\to\pi^-\pi^0\gamma\nu_\tau$ decays.}\label{AF:fig2}
\end{figure}
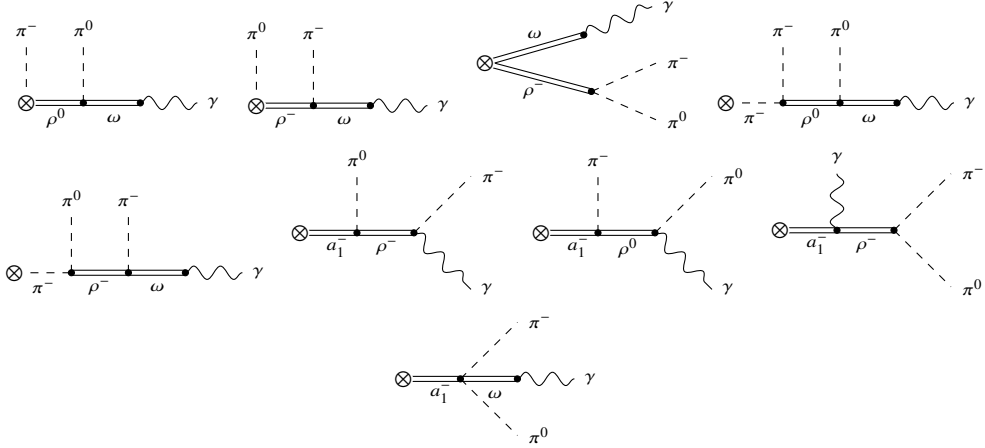

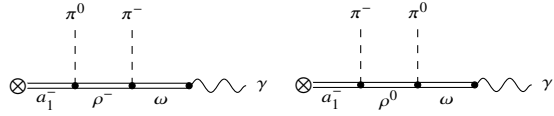
\begin{figure}[ht]
\begin{center}
\begin{tikzpicture}
	\draw[resonance] (-0.75,-4)--(1.5,-4);
	\draw[fill,white] (-0.75,-4) circle [radius=0.12];
	\node at (-0.75,-4) {\small$\otimes$};
	\node at (-0.375,-4.2) {\tiny$a_1^-$};
	\draw[scalar] (0,-4)--(0,-3.25) node[above] {\tiny$\pi^0$};
	\draw[fill] (0,-4) circle [radius=0.04];
	\node at (0.375,-4.2) {\tiny$\rho^-$};
	\node at (1.125,-4.2) {\tiny$\omega$};
	\draw[scalar] (0.75,-4)--(0.75,-3.25) node[above] {\tiny$\pi^-$};
	\draw[fill] (0.75,-4) circle [radius=0.04];
	\draw[gauge] (1.5,-4)--(2.25,-4) node[right] {\tiny$\gamma$};
	\draw[fill] (1.5,-4) circle [radius=0.04];
\end{tikzpicture}
\begin{tikzpicture}
	\draw[resonance] (3.5,-4)--(5.75,-4);
	\draw[fill,white] (3.5,-4) circle [radius=0.12];
	\node at (3.5,-4) {\small$\otimes$};
	\node at (3.875,-4.2) {\tiny$a_1^-$};
	\draw[scalar] (4.25,-4)--(4.25,-3.25) node[above] {\tiny$\pi^-$};
	\draw[fill] (4.25,-4) circle [radius=0.04];
	\node at (4.625,-4.2) {\tiny$\rho^0$};
	\node at (5.375,-4.2) {\tiny$\omega$};
	\draw[scalar] (5,-4)--(5,-3.25) node[above] {\tiny$\pi^0$};
	\draw[fill] (5,-4) circle [radius=0.04];
	\draw[gauge] (5.75,-4)--(6.5,-4) node[right] {\tiny$\gamma$};
	\draw[fill] (5.75,-4) circle [radius=0.04];
\end{tikzpicture}
\end{center}
\caption{Three-resonance exchange contributions from the $R\chi T$ to the axial-vector form factors of the $\tau^-\to\pi^-\pi^0\gamma\nu_\tau$ decays.}\label{AF:fig3}
\end{figure}

For the axial form factors, we have
\begin{subequations}\label{eq:AFF}\small\begin{align}
a_1&=a_1^0+a_1^R+a_1^{RR}+a_1^{RRR},\\
a_2&=a_2^0+a_2^R+a_2^{RR}+a_2^{RRR},\\
a_3&=a_3^R+a_3^{RR}+a_3^{RRR},\\
a_4&=a_4^R+a_4^{RR}+a_4^{RRR},
\end{align}\end{subequations}
where $a_i^{R}$, $a_i^{RR}$ and $a_i^{RRR}$ include up to $\mathcal{O}\left(p^6\right)$ contributions. The diagrams contributing to the Eq. (\ref{eq:AFF}) are shown in figures \ref{Ap:fig3}-\ref{AF:fig3}. These expressions can be found in App. D. in Ref. \cite{Miranda:2020wdg}.

\subsection{SD constraints}\label{subsec:SDC}
Including the complete set of operators \cite{Cirigliano:2006hb, Kampf:2011ty} that start contributing to the $\mathcal{O}(p^6)$ LECs, we have too many parameters which are allowed by the discrete symmetries of QCD and chiral symmetry that prevent us from making phenomenology predictions. For a detailed discussion you can see Section 1.5 in Ref. \cite{Miranda:2020wdg}. 

Imposing the asymptotic behavior of: the pion vector form factor, the $V-A$ correlator, the scalar form factor, and the $S-P$ correlator, the following constraints are found \cite{Ecker:1988te, Ecker:1989yg, Pich:2002xy, Weinberg:1967kj, Golterman:1999au, Jamin:2000wn, Jamin:2001zq}~\footnote{Other important relations are in App. \ref{App:SD}. }:
\begin{equation}\label{SD:eq01}\small\begin{split}
F_V G_V=F^2,&\qquad F_V^2-F_A^2=F^2,\\
F_V^2 M_V^2=F_A^2 M_A^2,&\qquad 4c_d c_m=F^2,\\
8\left(c_m^2-d_m^2\right)=F^2,&\qquad c_m=c_d=\sqrt{2}d_m=F/2.
\end{split}\end{equation}

The asymptotic behavior of the $2-$point Green function at $\mathcal{O}(p^4)$ predicts,
\begin{equation}\label{SDCEN}
F_V= \sqrt{2} F\,,\quad G_V = \frac{F}{\sqrt{2}}\,,\quad F_A=F\,,
\end{equation}
for the couplings of the $R\chi T$ Lagrangian \cite{Ecker:1989yg}. Conversely, the $2$- and $3$-point Green function up to $\mathcal{O}(p^6)$ \cite{Cirigliano:2006hb, Kampf:2011ty, Roig:2013baa}~\footnote{The contributions from operators with more than one resonance field are taken into account.} determine
\begin{equation}\label{SDCEN3}
F_V= \sqrt{3} F\,,\quad G_V = \frac{F}{\sqrt{3}}\,,\quad F_A=\sqrt{2}F\,.
\end{equation}
We will hereinafter refer to the constraints from the $2$- and $3$-point Green functions as `$F_V=\sqrt{2}F$' and `$F_V=\sqrt{3}F$', respectively. 

Since the $\kappa_{i}^{V}$ couplings are related with the $\omega$ exchange which is known to give an important contribution to the $\tau\to\pi\pi\gamma\nu_\tau$ decays, we perform a global fit using the relations for the resonance saturation of the anomalous sector LECs at NLO \cite{Kampf:2011ty}, the eqs. in App. \ref{App:SD} and the estimation of the LECs in \cite{Jiang:2015dba}. The fit outcomes are in App. \ref{Fit}.

\section{Decay observables}\label{sec:decayobs}
The differential decay width \cite{FloresTlalpa:2008zz} is given by 
\begin{equation}\label{Appx4:eq63}
d\Gamma =\frac{\lambda^{1/2}\left(s,m_{\pi^0}^2,m_{\pi^-}^2\right)}{2\left(4\pi\right)^6m_\tau^2 s}\overline{\left\vert\mathcal{M}\right\vert^2}\,dE_\gamma\,dx\,ds\,d\cos\theta_-\,d\phi_-,
\end{equation}
where $\overline{\left\vert\mathcal{M}\right\vert^2}$ is the unpolarized spin-averaged squared amplitude that corresponds to the $\tau^-\to\pi^-\pi^0\gamma\,\nu_\tau$ decays, and $E_\gamma$ is the photon energy in the $\tau$ rest frame. It is not worth to write down here the full analytical expression for $\overline{\left\vert\mathcal{M}\right\vert^2}$. The kinematics of this process can be found in App. \ref{App:kin}

\subsection{Decay spectrum}
Integrating the Eq. (\ref{Appx4:eq63}) over $E_\gamma$, $x$, $\cos\theta_{-}$ and $\phi_-$, we obtain the $\pi^-\pi^0$ hadronic invariant distribution. Since the decay spectrum is IR divergent due to soft photons, we require to introduce a photon energy cut, $E_\gamma^{cut}$, which is related to the experimental resolution. 

\begin{figure}[!ht]
\includegraphics[width=10cm]{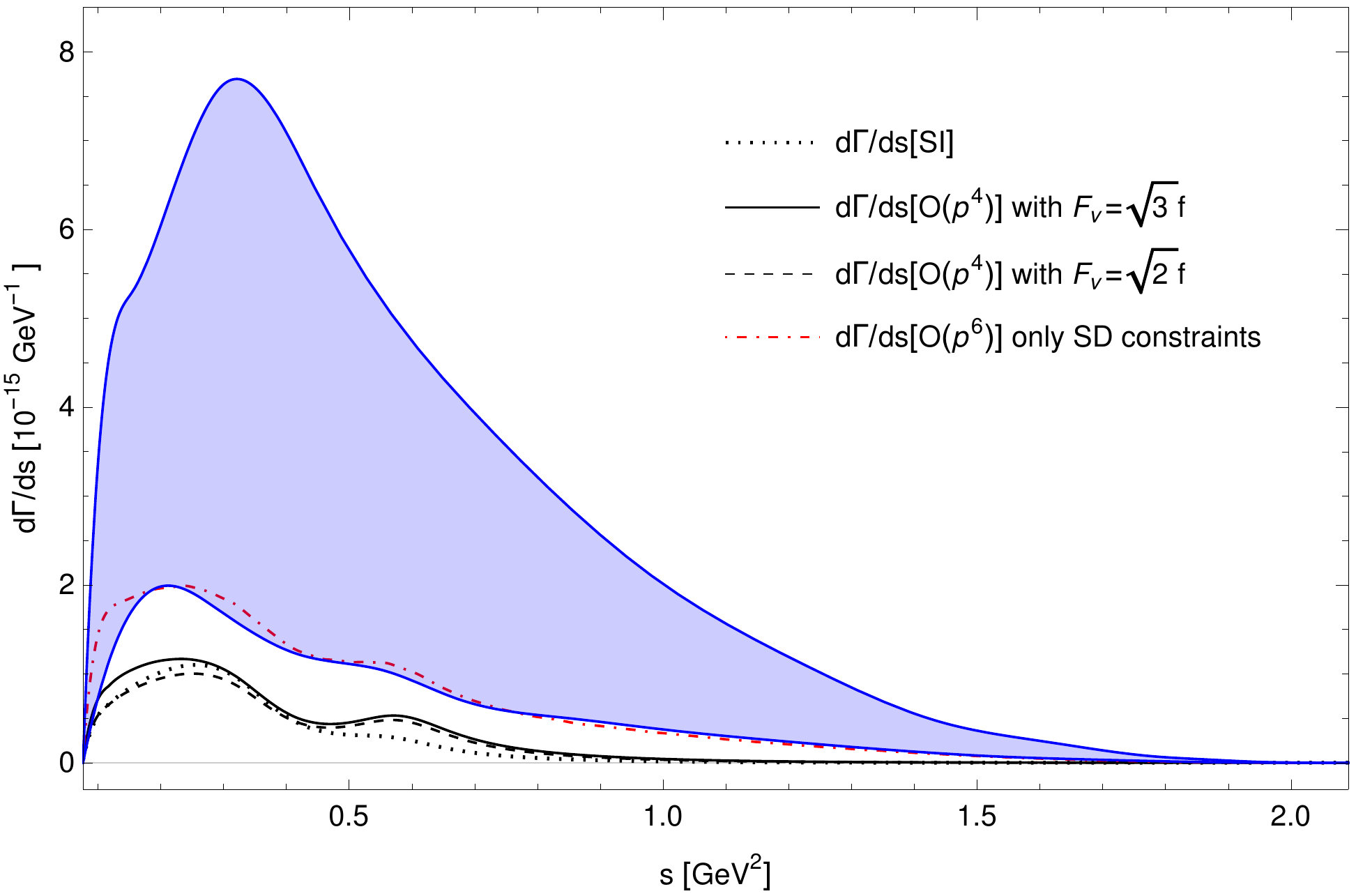}		
\centering			
\caption{The $\pi^-\pi^0$ hadronic invariant mass distributions for $E_\gamma^{cut}=300\,\text{MeV}$. The solid and dashed lines represent the $\mathcal{O}\left(p^4\right)$ corrections using $F_V=\sqrt{3}F$ and $F_V=\sqrt{2}F$, respectively. The dotted line represents the Bremsstrahlung contribution (SI). The dotdashed red line corresponds to using only SD constraints at $\mathcal{O}\left(p^6\right)$ and the blue shaded region overestimates the corresponding uncertainties.}\label{Appx4:fig5}
\end{figure}

In figure \ref{Appx4:fig5}, we can see the prediction for the decay spectrum for $E_\gamma^{cut}=300\,\text{MeV}$. The dotted line indicates the limit where all the structure-dependent form factors vanish, i.e. $v_i=a_i=0$. The predictions at $\mathcal{O}(p^4)$ using $F_V=\sqrt{2}F$ and $F_V=\sqrt{3}F$, which are discussed in Sec. \ref{subsec:SDC}, are denoted by the dashed and solid line, respectively. The dotdashed red line corresponds to taking the limit where all the couplings at $\mathcal{O}(p^6)$ vanish except for those constrained by SD and the band overestimates the corresponding uncertainties. Including the $\mathcal{O}(p^6)$ corrections, the decay spectrum receives a noticeable enhancement at low $s$.

\subsection{Branching ratio}
\begin{figure}[!ht]
\includegraphics[width=10cm]{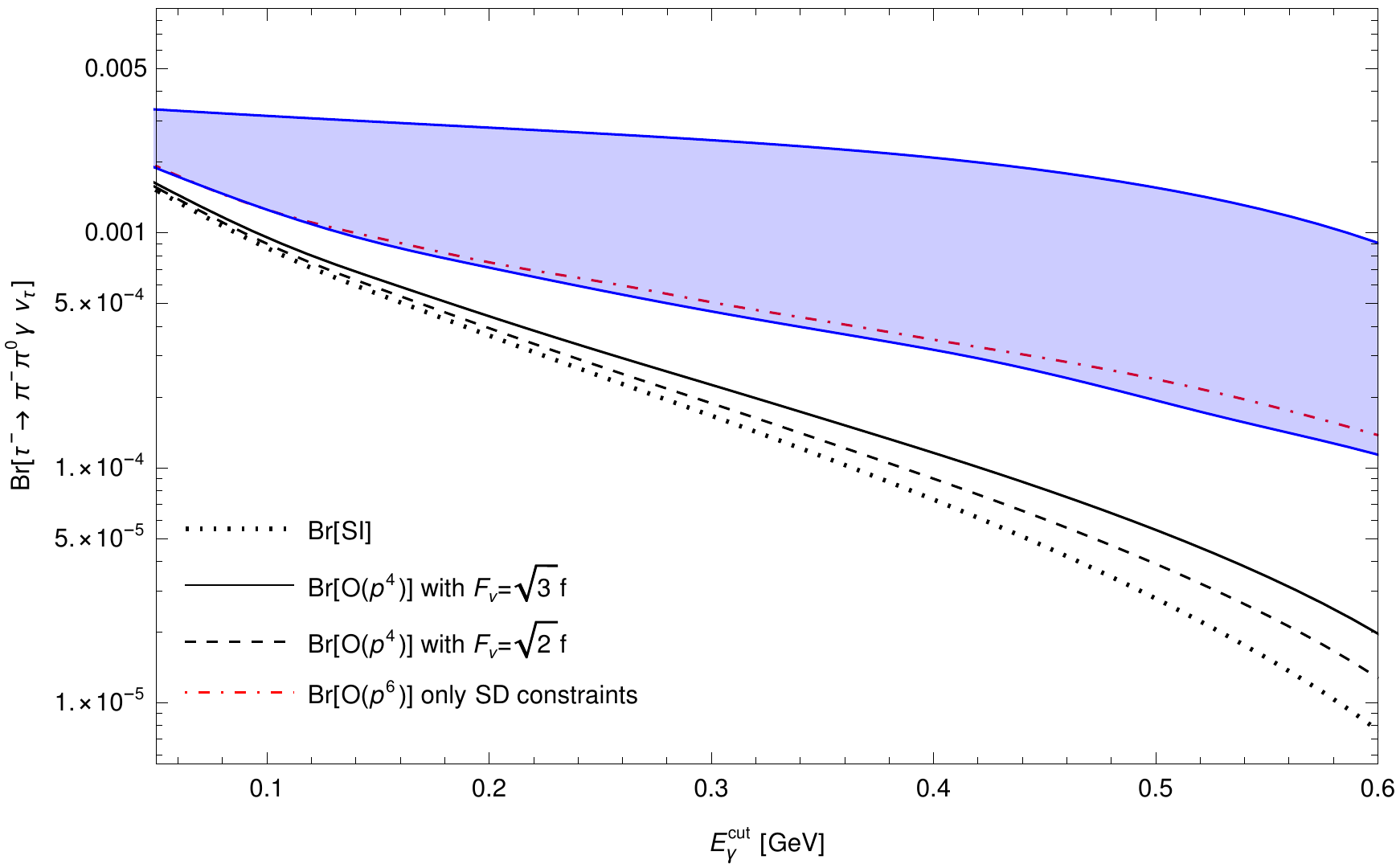}	
\centering			
\caption{Branching ratio for the $\tau^-\to\pi^-\pi^0\gamma\nu_\tau$ decays as a function of $E^{cut}_\gamma$.  The dotted line represents the Bremsstrahlung contribution, the solid and dashed lines represent the $\mathcal{O}\left(p^4\right)$ corrections using $F_V=\sqrt{3}F$ and $F_V=\sqrt{2}F$, respectively. The dotdashed red line is the $\mathcal{O}\left(p^6\right)$ contribution using only SD constraints and neglecting all other couplings. The blue shaded region overestimates the $\mathcal{O}\left(p^6\right)$ uncertainties.}\label{Appx4:fig6}
\end{figure}

By integrating upon the $s$ variable the decay spectrum, we obtain the branching ratio for the $\tau\to\pi\pi\gamma\nu_\tau$ decays. Figure \ref{Appx4:fig6} shows the prediction for the branching ratio as a function of $E_\gamma^{cut}$ obtained using the different orders of approximation for the structure-dependent terms in Eqs. (\ref{FFW:V}) and (\ref{FFW:A}).

According to figs. \ref{Appx4:fig5} and \ref{Appx4:fig6}, measurements of the $\pi\pi$ invariant mass and the partial decay width, for a reasonable cut on $E_\gamma$ (at low enough energies the inner bremmstrahlung contribution hides completely any structure-dependent effect), could decrease substantially the uncertainty of the $\mathcal{O}\left(p^6\right)$ computation~\footnote{The photon spectrum measurement would also help to this task \cite{Miranda:2020wdg}.}. This was already emphasized in Ref. \cite{Cirigliano:2002pv} but remained unmeasured at BaBar and Belle. We hope these data can finally be acquired and analyzed at Belle-II.

\section{Hadronic Vacuum Polarization}\label{sec:HVP}
The leading contributions to the hadronic vacuum polarization (HVP) employing a dispersion relation \cite{Gourdin:1969dm}, which complies with unitarity and analyticity, are given by
\begin{equation}
a^{HVP,LO}_\mu=\frac{1}{4\pi^3}\int^{\infty}_{s_{thr}} ds \,K(s) \sigma^0_{e^- e^+\to hadrons}(s),
\end{equation}
where $K(s)$ is a smooth QED kernel \cite{Brodsky:1967sr} concentrated at low energies, which increases the $E\lesssim M_\rho$ contribution, and $ \sigma_{e^- e^+\to hadrons}^0(s)$ is the bare hadronic cross section. We can relate the hadronic spectral function from $\tau$ decays to the $e^+e^-$ hadronic cross section by including the radiative corrections and the IB  effects. 

For the $\pi\pi(\gamma)$ final state, we have \cite{Cirigliano:2001er, Cirigliano:2002pv}:
\begin{equation}
\sigma_{\pi\pi(\gamma)}^0=\left[\frac{K_\sigma(s)}{K_\Gamma (s)}\frac{d\Gamma_{\pi\pi[\gamma]}}{ds}\right]\frac{R_{IB}(s)}{S_{EW}},
\end{equation}
where
\begin{equation}\begin{split}
K_\Gamma(s)&=\frac{G_F^2\vert V_{ud}\vert^2 m_\tau^3}{384 \pi^3}\left(1-\frac{s}{m_\tau^2}\right)^2\left(1+\frac{2s}{m_\tau^2}\right),\quad K_\sigma(s)=\frac{\pi \alpha^2}{3s},
\end{split}\end{equation}
and the IB corrections 
\begin{equation}
R_{IB}(s)=\frac{FSR(s)}{G_{EM}(s)}\frac{\beta^3_{\pi^+\pi^-}}{\beta^3_{\pi^+\pi^0}}\left\vert\frac{F_V(s)}{f_+(s)}\right\vert^2.
\end{equation}
The $S_{EW}$ term encodes the SD electroweak corrections \cite{Sirlin:1974ni, Sirlin:1977sv, Sirlin:1981ie, Marciano:1985pd, Marciano:1988vm, Braaten:1990ef, Marciano:1993sh, Erler:2002mv} and $FSR(s)$ accounts for the final-state radiation from pions \cite{Schwinger:1989ix, Drees:1990te}. The $\beta^3_{\pi^+\pi^-}/\beta^3_{\pi^+\pi^0}$ term is a phase space factor and the last term in $R_{IB}(s)$ is a ratio between the neutral ($F_V(s)$) and the charged ($f_+(s)$) pion form factor.

\begin{figure}[!ht]
\includegraphics[width=7.4cm]{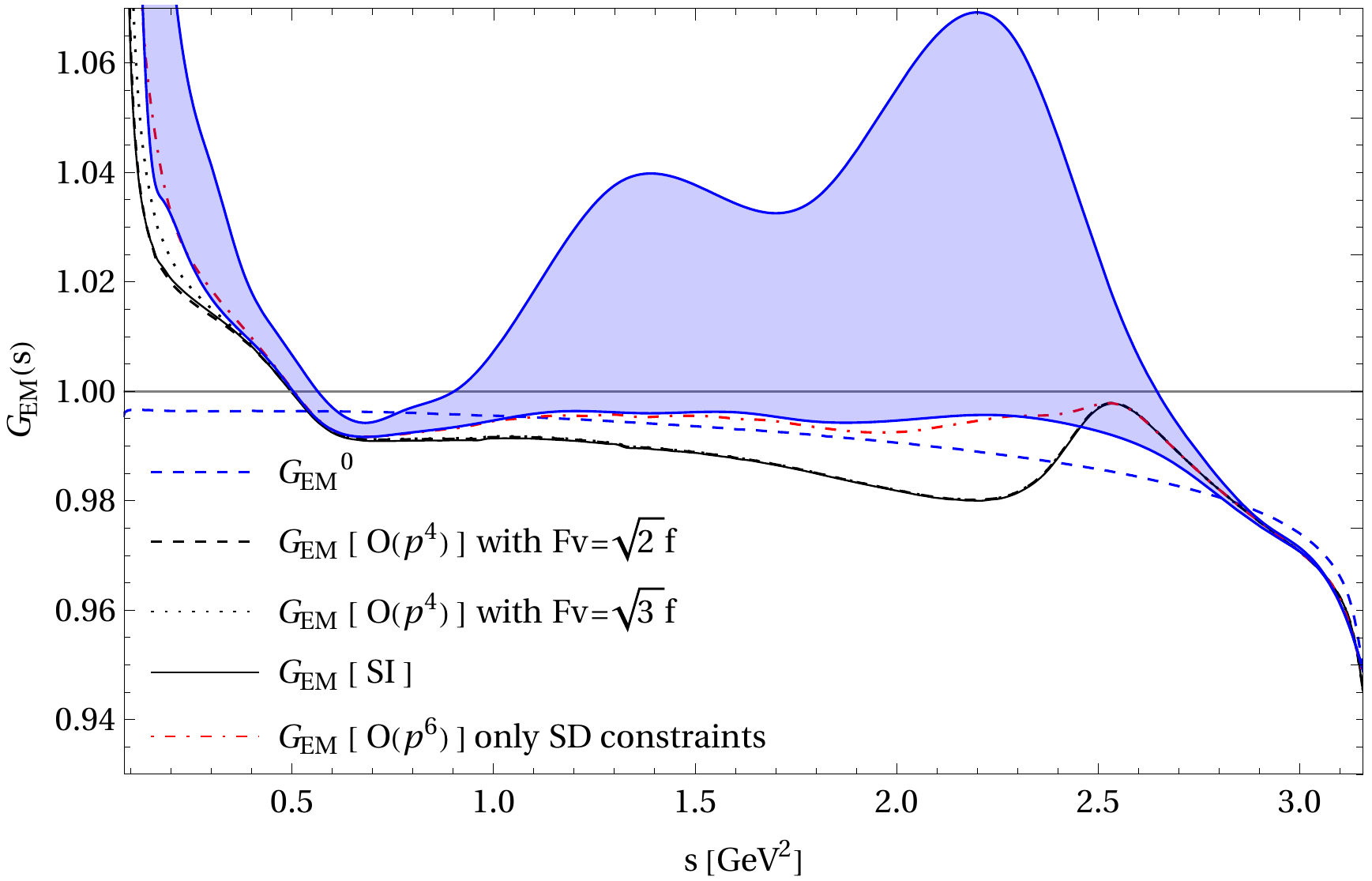}	
\includegraphics[width=7.4cm]{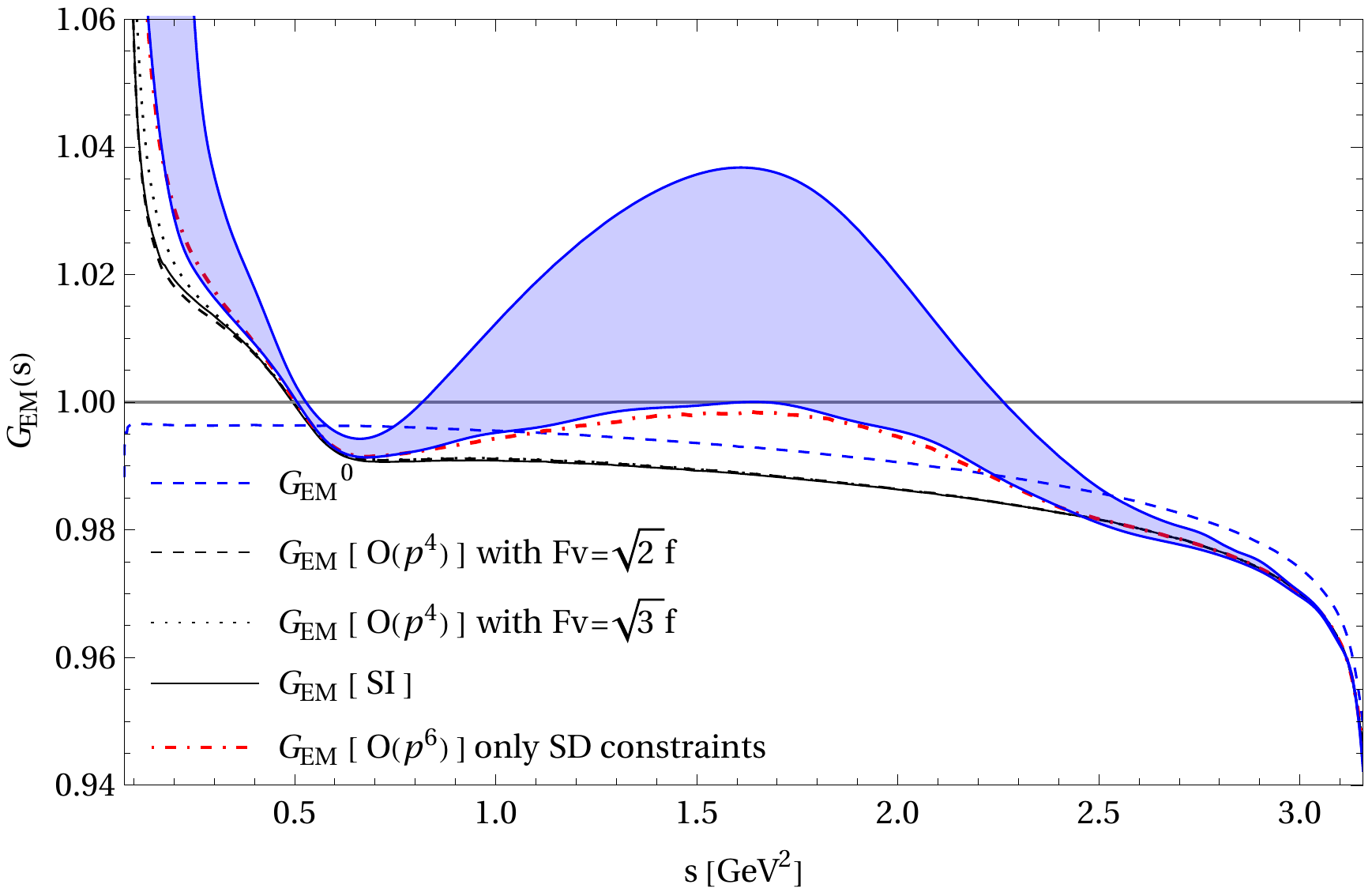}	
\centering			
\caption{Correction function $G_{EM}^{(0)}\left(s\right)$ (blue dashed line). The solid line shows the $G_{EM}(s)$ function neglecting the structure-dependent part (SI), i.e. by taking $v_i=a_i=0$, the dashed and dotted lines are the $\mathcal{O}\left(p^4\right)$ $G_{EM}(s)$ function  (with either $F_V=\sqrt{2}F$ or $F_V=\sqrt{3}F$ constraints). The blue shaded region is the full $\mathcal{O}\left(p^6\right)$ contribution, including (overestimated) uncertainties. The left-hand side plot corresponds to the dispersive parametrization \cite{GomezDumm:2013sib} while the right-hand side corresponds to the Guerrero-Pich parametrization \cite{Guerrero:1997ku} of the form factor (the latter was used in Ref. \cite{Cirigliano:2002pv}).}\label{Appx4:fig3}
\end{figure}

The $G_{EM}(s)$ function is obtained by adding up the contributions due to virtual and real photons and integrating over the $u\equiv(P-p_{-})^2$ variable,
\begin{equation}\label{Appx4:eq49}\begin{split}
\left.\frac{d\Gamma}{ds}\right\vert_{\pi\pi(\gamma)}=&\frac{G_F^2 \vert V_{ud}\vert^2 m_\tau^3 S_{EW}}{384\pi^3}\left\vert f_+(s)\right\vert^2 \left(1-\frac{s}{m_\tau^2}\right)^2\left(1-\frac{4m_\pi^2}{s}\right)^{3/2}\left(1+\frac{2s}{m_\tau^2}\right)G_{EM}(s).
\end{split}\end{equation}
The results are shown in figure \ref{Appx4:fig3} for the different approximations of the $G_{EM}(s)$ function using the dispersive (left-hand) and the exponential (right-handed) representation of the pion form factor. The $G_{EM}^{0}(s)$ contribution was obtained using the leading Low approximation in Eq. (\ref{eq:Lowapp}).

We can estimate the effect of each IB correction through $\Delta a_\mu^{HVP,LO}[\pi\pi]$ \cite{Cirigliano:2002pv},
\begin{equation}\label{IB:eq69}
\Delta a_\mu^{HVP,LO}=\frac{1}{4\pi^3}\int_{s_1}^{s_2}ds\,K(s)\left[\frac{K_\sigma(s)}{K_\Gamma (s)}\frac{d\Gamma_{\pi\pi[\gamma]}}{ds}\right]\left(\frac{R_{IB}(s)}{S_{EW}}-1\right),
\end{equation}
which measures the departure from the isospin-limit, i.e. $R_{IB}(s)=1$ and $S_{EW}=1$. It is challenging to evaluate the corrections owing to the ratio of the form factors. For this enterprise, we have followed two alternatives, dubbed FF1 and FF2:

\begin{itemize}
\item In FF1, we use for the $\rho-\omega$ mixing parameter $\theta_{\rho\omega}=(-3.5\pm0.7)\times10^{-3}\:\mathrm{GeV}^2$ \cite{Cirigliano:2002pv} and $\Gamma_{\rho^0}-\Gamma_{\rho^+}=0.3\pm1.3$ MeV, $m_{\rho^\pm}-m_{\rho^0}=0.7\pm0.8$ MeV and $m_{\rho^0}=775.26\pm0.25$ MeV from PDG \cite{ParticleDataGroup:2020ssz}. 
\item In FF2, we use the same numerical input as in FF1 except by the rho width, which is $\Gamma_{\rho^0\to\pi^+\pi^-\gamma}-\Gamma_{\rho^\pm\to\pi^\pm\pi^0\gamma}=0.45\pm 0.45\,\text{MeV}$ \cite{Cirigliano:2002pv}.
\end{itemize}

The outcomes are summarized in Table \ref{HVP:tab3} using DR form factor. The results obtained for the $G_{EM}^{(0)}$ and the complete $\mathcal{O}\left(p^4\right)$ contribution (with $F_V=\sqrt{2}F$) agree with those in Ref. \cite{Cirigliano:2002pv}. The uncertainties at $\mathcal{O}(p^4)$ were obtained using the dashed and dotdashed red line in fig. \ref{Appx4:fig3}. On the other hand, the errors at $\mathcal{O}(p^6)$ were estimated using the blue region in the same plot.

\begin{table}[htbp]
\begin{center}
\small{\begin{tabular}{|c|c|c|c|c|c|c|c|}
\hline
$\left[s_1,s_2\right]$ & $\mathrm{S_{EW}}$ & $\mathrm{PS}$ & $ \mathrm{FSR}$ & $\mathrm{FF1}$ &  $\mathrm{FF2}$ & $\mathrm{EM}$ & $\mathrm{EM}$ \\
 &  &  &  &  &  & $\,\mathcal{O}(p^4)$ & $\,\mathcal{O}(p^6)$\\
\hline
\hline
$\left[4m_\pi^2,1\,\mathrm{GeV}^2\right]$ & $-101.1$ & $-74.1$ & $+44.7$ & $+41.8\pm49.0 $  & $+78.4\pm24.5$ & $-17.0^{+5.7}_{-15.4}$ & $-74.8^{+63.5}_{-44.0}$ \\
$\left[4m_\pi^2,2\,\mathrm{GeV}^2\right]$ & $-103.1$ & $-74.4$ & $+45.5$ & $+40.9\pm48.9$  & $+77.6\pm24.0$ & $-16.0^{+5.7}_{-15.9}$ & $-75.9^{+65.6}_{-45.5}$ \\
$\left[4m_\pi^2,3\,\mathrm{GeV}^2\right]$ & $-103.1$ & $-74.5$  & $+45.5$ & $+40.9\pm48.9$ & $+77.6\pm24.0$ & $-15.9^{+5.7}_{-16.0}$ & $-75.9^{+65.7}_{-44.6}$ \\
$\left[4m_\pi^2,m_\tau^2\right]$ & $-103.1$ & $-74.5$  & $+45.5$ & $+40.9\pm48.9$ & $+77.6\pm24.0$ & $-15.9^{+5.7}_{-16.0}$ & $-75.9^{+65.7}_{-45.6}$ \\
\hline
\end{tabular}}
\caption{Contributions to $\Delta a_\mu^{HVP,LO}$ in units of $10^{-11}$ using the DR form factor as the reference one.}
\label{HVP:tab3}
\end{center}
\end{table}

An important cross-check is the branching fraction $B_{\pi\pi^0}=\Gamma(\tau\to\pi\pi^0\nu_\tau)/\Gamma_\tau$ which can be directly measured in experiments. It can also be evaluated from the isovector component of the $e^+e^-\to\pi^+\pi^-(\gamma)$ cross section after taking into account the IB corrections. The branching fraction is given by
\begin{equation}
B_{\pi\pi^0}^{CVC}= B_{e}\int_{4m_\pi^2}^{m_\tau^2}ds\,\sigma_{\pi^+\pi^-(\gamma)}(s)\mathcal{N}(s)\frac{S_{EW}}{R_{IB}(s)},
\end{equation}
where 
\begin{equation}
\mathcal{N}(s)=\frac{3\left\vert V_{ud}\right\vert^2}{2\pi\alpha_0^2 m_\tau^2}s\left(1-\frac{s}{m_\tau^2}\right)^2\left(1+\frac{2s}{m_\tau^2}\right).
\end{equation}
Using the most recent data obtained from BaBar \cite{Lees:2012cj}~\footnote{We thank to Alex Keshavarzi and Bogdan Malaescu for providing us tables with the measurement of the $e^+e^-\to\pi^+\pi^-(\gamma)$ cross section.} for the $e^+e^-\to\pi^+\pi^-(\gamma)$ cross section, we obtain
\begin{equation}
B_{\pi\pi^0}^{CVC}=(24.68  \pm0.11    \pm0.10  \pm0.01  \pm 0.01 \pm 0.02^{+0.03}_{-0.00})\,\%,  \text{ at }\mathcal{O}(p^4),
\end{equation}
and
\begin{equation}
B_{\pi\pi^0}^{CVC}=(24.70  \pm0.11    \pm0.10  \pm0.01  \pm 0.01 \pm 0.02^{+0.21}_{-0.01})\,\%,  \text{ at }\mathcal{O}(p^6),
\end{equation}
where the first error corresponds to the statistical experimental uncertainty on $\sigma_{\pi\pi(\gamma)}$, the second is related to uncertainty on the $\rho^+-\rho^0$ width difference, the third to the uncertainty in the $\rho^+-\rho^0$ mass difference, the fourth to the uncertainty of the $\rho-\omega$ mixing and the fifth corresponds to the corrections induced by FSR on $B_{\pi\pi^0}^{CVC}$, which reduces $\sim-0.20(2)\%$ the $\pi\pi$ branching fraction. The last error is related to the $G_{EM}(s)$ function.

These results are in good agreement with the value reported by the Belle \cite{Fujikawa:2008ma} collaboration, $B_{\pi\pi^0}^{\tau}=(25.24\pm0.01\pm0.39)\%
$, where the first uncertainty is statistical and the second is systematic. Nonetheless, they are in some tension with the very precise ALEPH measurement $B_{\pi\pi^0}^{\tau}=(25.471\pm0.097\pm0.085)\%$ \cite{Schael:2005am}.

\begin{figure}[htbp]
\includegraphics[width=7.4cm]{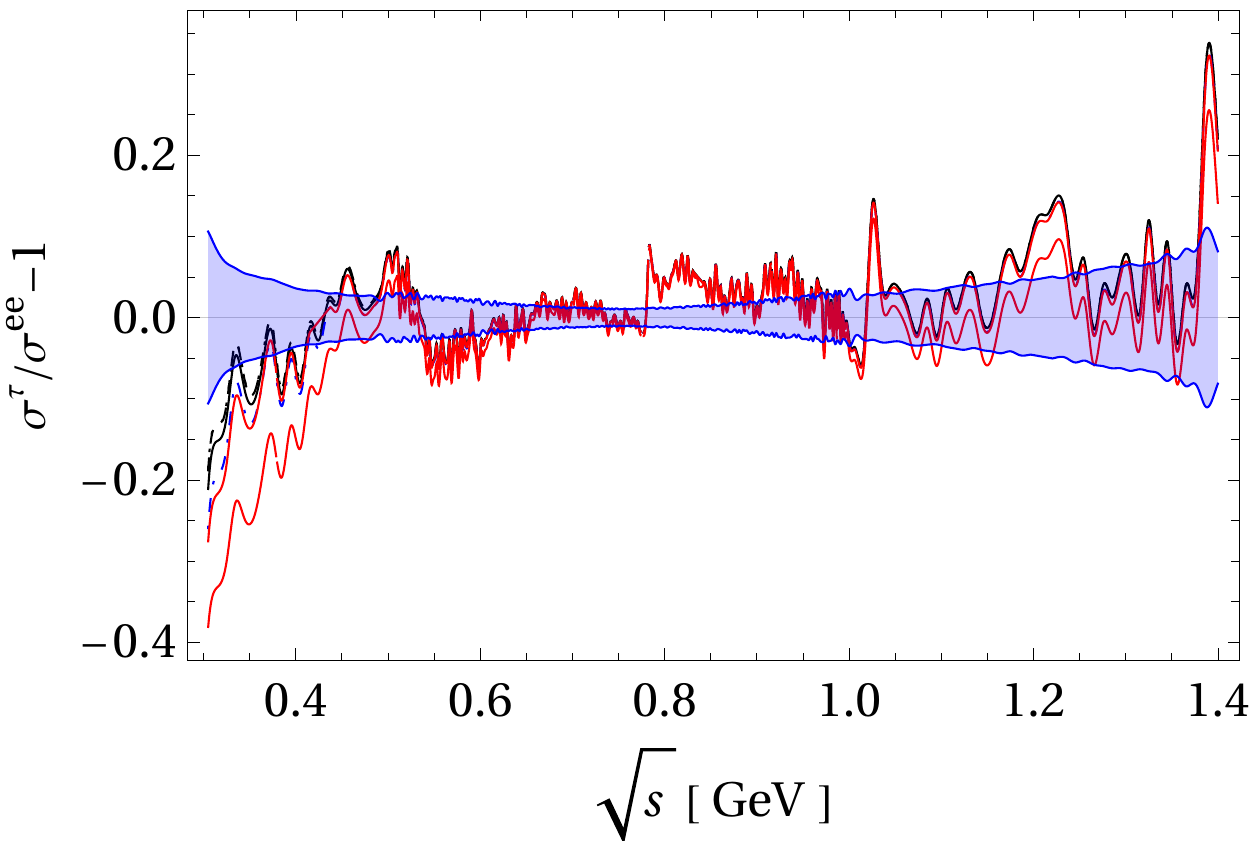}	
\includegraphics[width=7.4cm]{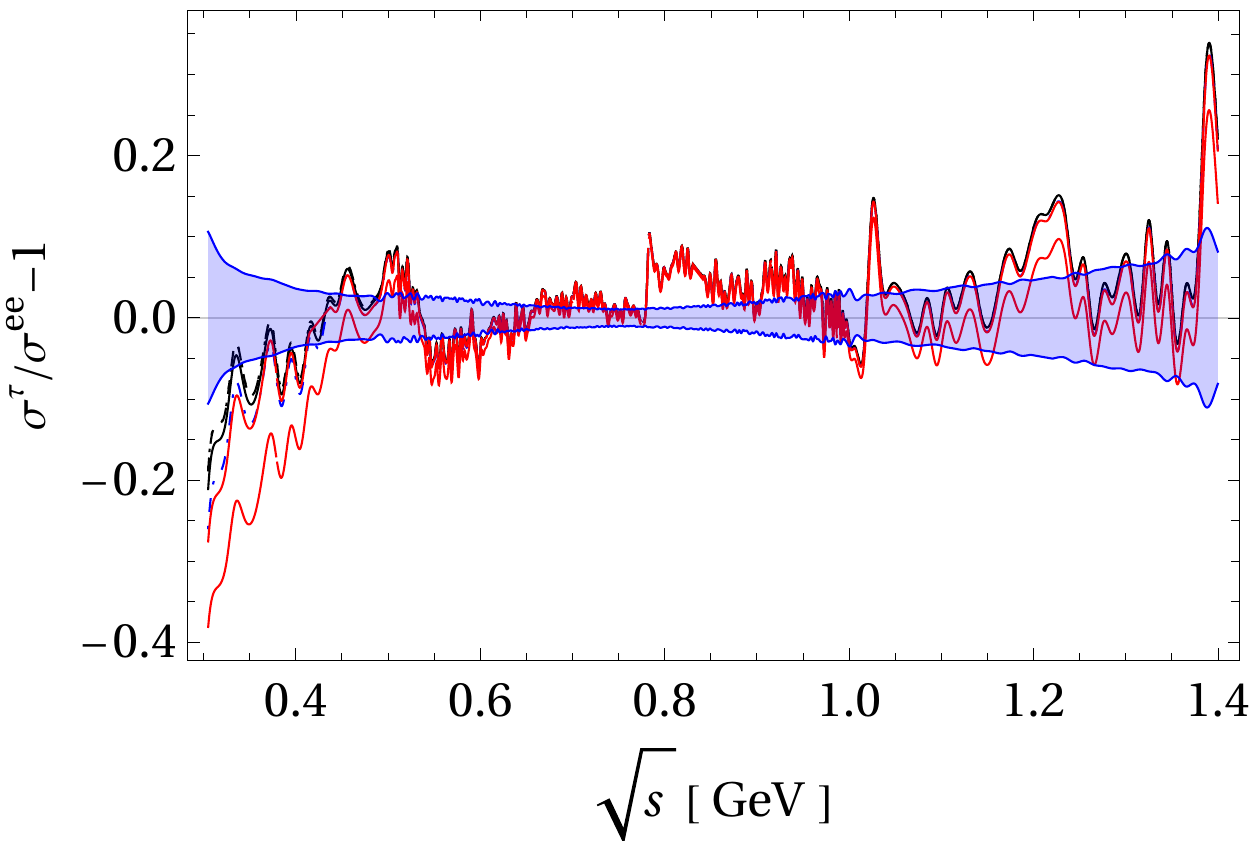}	
\includegraphics[width=7.4cm]{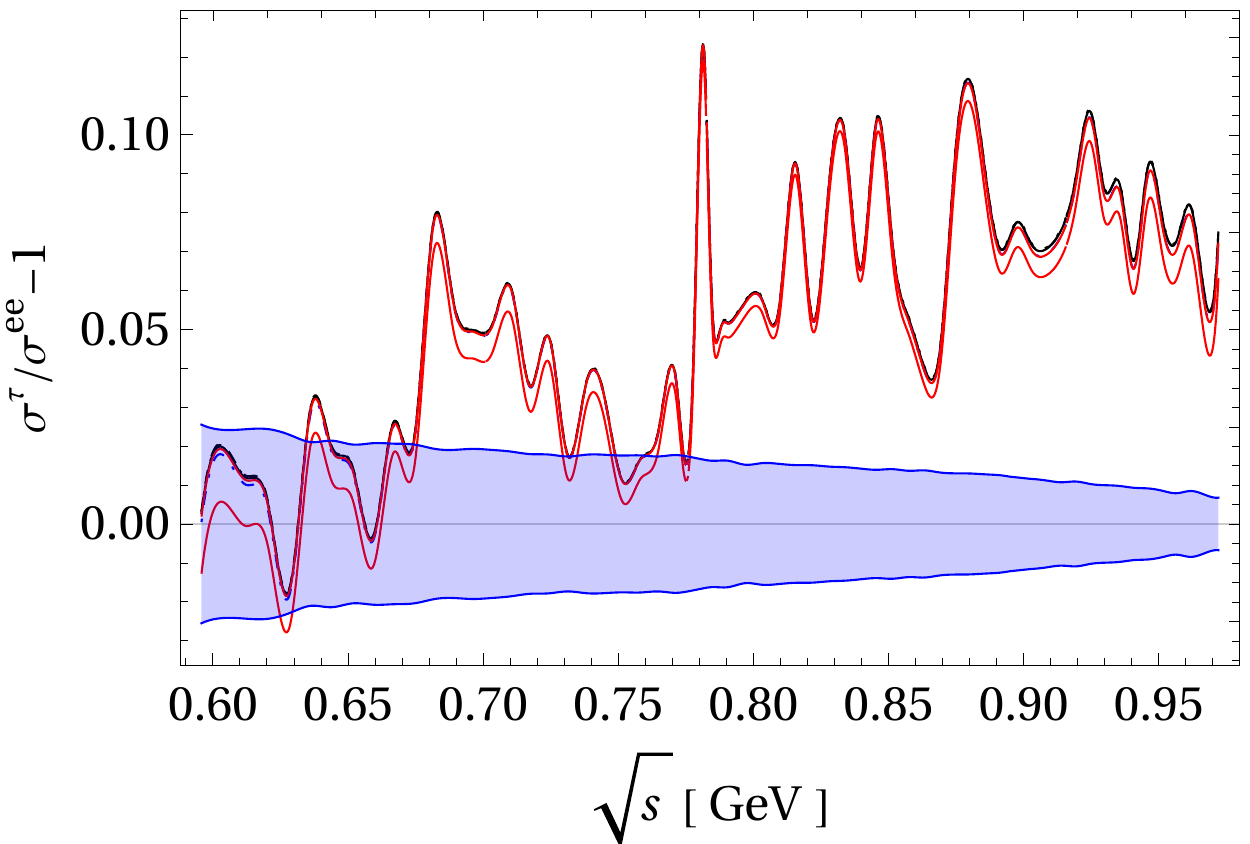}	
\includegraphics[width=7.4cm]{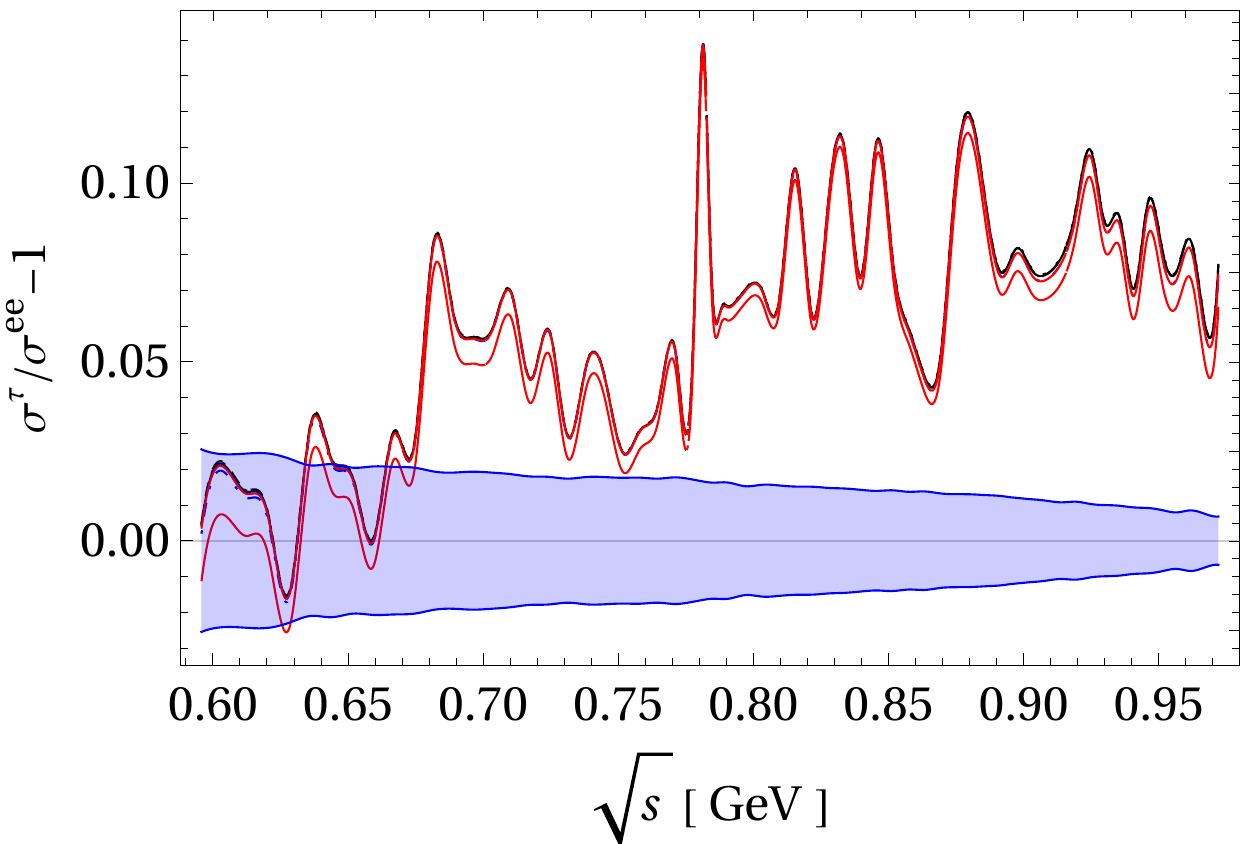}	
\centering			
\caption{Comparison between the different data sets from BaBar (above) and KLOE (below) with $\Delta\Gamma_{\pi\pi\gamma}=1.5\,\mathrm{MeV}$ (left-hand) and $\Delta\Gamma_{\pi\pi\gamma}=0.45\,\mathrm{MeV}$ (right-hand) for FF1 and FF2, respectively. The blue region corresponds to the experimental error on $\sigma_{\pi\pi(\gamma)}$. The solid and dashed lines represent the contributions with $F_V=\sqrt{3}F$ and $F_V=\sqrt{2}F$ at $\mathcal{O}(p^4)$, respectively. The dotted line is the SI contribution. The red line depicts the envelope of $G_{EM}(s)$ at $\mathcal{O}(p^6)$, that overestimates the uncertainty at this order. The blue dotdashed line is the $\mathcal{O}(p^6)$ contribution using only SD constraints.}\label{ee:fig1}
\end{figure}

We show in fig. \ref{ee:fig1} the prediction for the $e^+e^-\to\pi^+\pi^-$ cross section using the data reported by Belle \cite{Fujikawa:2008ma} (as it is the most precise measurement of this spectrum) for the normalized spectrum $(1/N_{\pi\pi})(dN_{\pi\pi}/ds)$ compared to the last measurements from BaBar \cite{Lees:2012cj} and KLOE \cite{Babusci:2012rp}~\footnote{We have chosen to show in the comparison these two $e^+e^-$ data sets as the results from both Colls. are those deviating the most, and thus mainly responsible from the tension in $\sigma(e^+e^-\to\pi^+\pi^-)$.}.

In fig. \ref{ee:fig1} the $\tau$-based prediction is obtained using the $\mathcal{O}(p^4)$ result for $G_{EM}(s)$, with the estimated uncertainty from missing higher-order corrections given by the result at $\mathcal{O}(p^6)$ (employing only the SD constraints). The blue dotdashed line shown overestimates the error at $\mathcal{O}(p^6)$.

From fig. \ref{ee:fig1}, we observe good agreement between the BaBar data and the $\tau$ decays prediction (slightly better for FF1). The previous comparisons make us consider our evaluation with FF1 the reference one (so that its difference with FF2 will assess the size of the error induced by IB among the $\rho\to\pi\pi\gamma$ decay channels)~\footnote{We, nevertheless, recall that recent BESIII data \cite{BESIII:2015equ} and evaluations within the Hidden Local Symmetry model \cite{Benayoun:2011mm, Benayoun:2012wc, Benayoun:2015gxa, Benayoun:2019zwh} agree better with the KLOE data than with BaBar's.}.

Taking into account all di-pion tau decay data from the ALEPH \cite{Schael:2005am}, Belle \cite{Fujikawa:2008ma}, CLEO \cite{Anderson:1999ui} and OPAL \cite{Ackerstaff:1998yj} Colls., we get the combined tau-data contribution
\begin{equation}\label{pipifromtau_p4}
 10^{10}\cdot a_\mu^{HVP,LO|_{\pi\pi,\tau\;\mathrm{data}}}\,=\,519.6\pm{2.8_{\mathrm{spectra+BRs}}}{^{+1.9}_{-2.1}}_{\mathrm{IB}}\,,
\end{equation}
at $\mathcal{O}(p^4)$ and 
\begin{equation}\label{pipifromtau_p6}
10^{10}\cdot a_\mu^{HVP,LO|_{\pi\pi,\tau\;\mathrm{data}}}\,=\,514.6\pm{2.8_{\mathrm{spectra+BRs}}}{^{+5.0}_{-3.9}}_{\mathrm{IB}}\,,
\end{equation}
at $\mathcal{O}(p^6)$.

When eqs.~(\ref{pipifromtau_p4}) and (\ref{pipifromtau_p6}) are supplemented with the four-pion tau decays measurements (up to $1.5$ GeV) and with $e^+e^-$ data at larger energies in these modes (and with $e^+e^-$ data in all other channels making up the hadronic cross section), we get \cite{Davier:2019can,Davier:2013sfa}

\begin{equation}\label{HVP,LO tau p4}
10^{10}\,\cdot\, a_\mu^{HVP,LO|_{\tau\;\mathrm{data}}}\,=\,705.7\pm{2.8_{\mathrm{spectra+BRs}}}{^{+1.9}_{-2.1}}_{\mathrm{IB}}\pm2.0_{\mathrm{e^+e^-}}\pm0.1_{\mathrm{narrow\,res}}\pm0.7_{\mathrm{QCD}}\,,
\end{equation}
at $\mathcal{O}(p^4)$, and 

\begin{equation}\label{HVP,LO tau p6}
10^{10}\,\cdot\, a_\mu^{HVP,LO|_{\tau\;\mathrm{data}}}\,=\,700.7\pm{2.8_{\mathrm{spectra+BRs}}}{^{+5.0}_{-3.9}}_{\mathrm{IB}}\pm2.0_{\mathrm{e^+e^-}}\pm0.1_{\mathrm{narrow\,res}}\pm0.7_{\mathrm{QCD}}\,,
\end{equation}

at $\mathcal{O}(p^6)$ and we have also included the uncertainties corresponding to using $e^+e^-$ data for those contributions not covered by tau decay measurements and to the inclusion of narrow resonances and the perturbative QCD part.

When all other (QED, EW and subleading hadronic) contributions are added to eqs. (\ref{HVP,LO tau p4}) and (\ref{HVP,LO tau p6}) according to Ref. \cite{Aoyama:2020ynm}, the $4.2\sigma$ \cite{Aoyama:2020ynm} deficit of the SM prediction with respect to the experimental average (FNAL+BNL) \cite{Muong-2:2021ojo,Bennett:2006fi} is reduced to

\begin{equation}
 \Delta a_\mu \equiv a_\mu^{exp}-a_\mu^{SM}=(12.5\pm 6.0)\cdot 10^{-10}\,,
\end{equation}
at $\mathcal{O}(p^4)$, and 

\begin{equation}
\Delta a_\mu \equiv a_\mu^{exp}-a_\mu^{SM}=(17.5^{+6.8}_{-7.5})\cdot 10^{-10}\,,
\end{equation}
at $\mathcal{O}(p^6)$, which are $2.1$ and $2.3\,\sigma$, respectively.

\begin{figure}[ht]
    \centering
    \includegraphics[width=13.3cm]{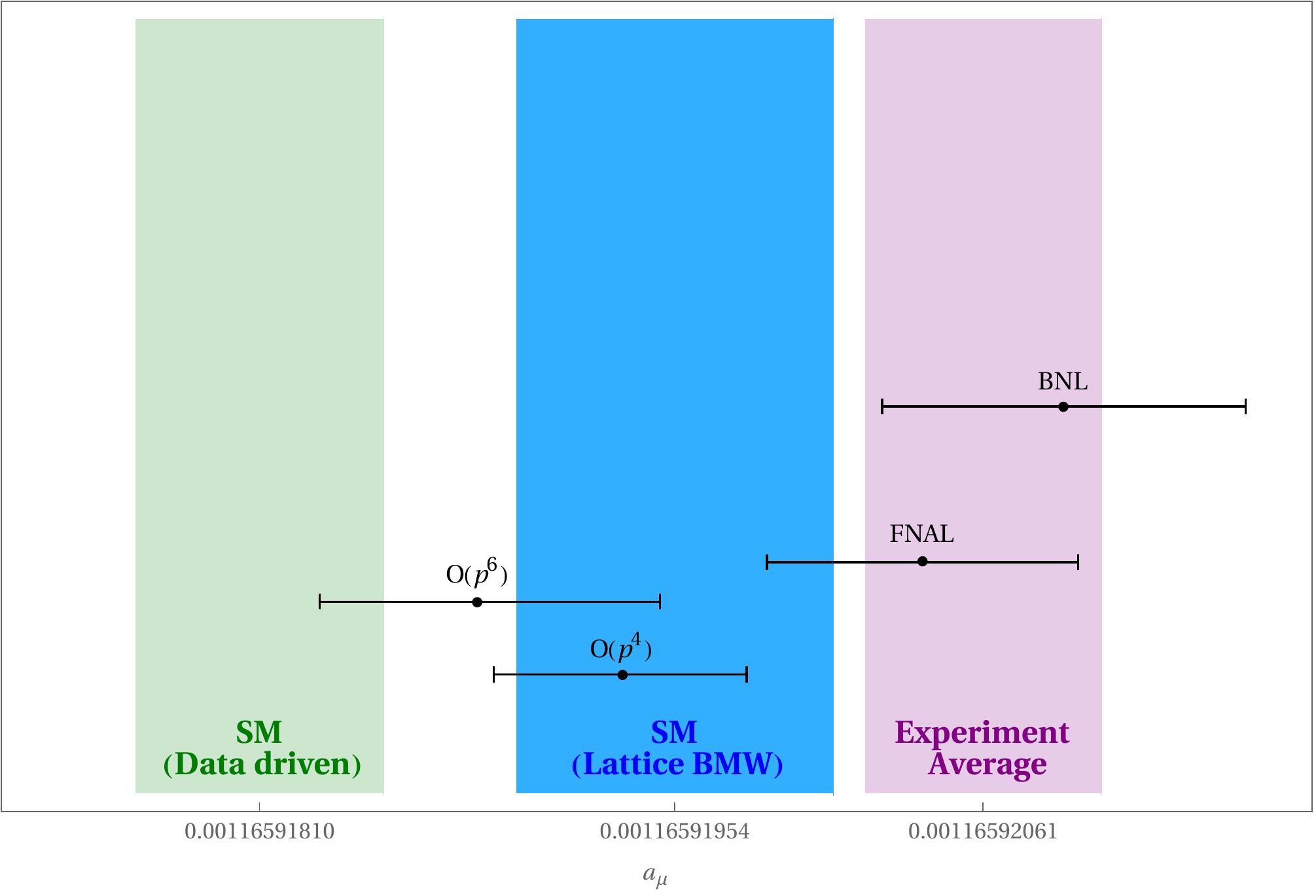}
    \caption{Comparison between the experimental values of $a_\mu$ from BNL \cite{Bennett:2006fi} and FNAL \cite{Muong-2:2021ojo} with respect to the Muon $g-2$ Theory Initiative recommended value \cite{Aoyama:2020ynm}, the lattice QCD calculation from the BMW collaboration \cite{Borsanyi:2020mff} and our results \cite{Miranda:2020wdg}.}
    \label{fig:amu_us}
\end{figure}

In figure \ref{fig:amu_us} we show a comparison between our $\mathcal{O}(p^4)$ and $\mathcal{O}(p^6)$ calculation with respect to the estimation based in the $e^+e^-$ data driven \cite{Aoyama:2020ynm} and the lattice results from the BMW collaboration \cite{Borsanyi:2020mff}.

\section{Conclusions}\label{sec:conc}
There is a global effort in improving the hadronic contributions to $a_\mu$. Specifically, dedicated studies to improve the HVP part from lattice, dispersion relations and improved $e^+e^-$ data and Monte Carlos are being undertaken.

The observables for the $\tau\to\pi\pi\gamma\nu_\tau$ decays have the potential to reduce drastically the errors in our estimation.

Our IB corrections improve the agreement between $e^+e^-$ and tau data, on the spectrum and the branching ratio.

Evaluating the HVP, LO contributions from tau data, we get  $a_\mu^{HVP,LO|_{\tau\;\mathrm{data}}}\,=\,(705.7^{+4.0}_{-4.1})\cdot10^{-10}$ at $\mathcal{O}(p^4)$, and $a_\mu^{HVP,LO|_{\tau\;\mathrm{data}}}\,=\,(700.7^{+6.1}_{-5.2})\cdot10^{-10}$ at $\mathcal{O}(p^6)$. This reduces the anomaly $\Delta a_\mu \equiv a_\mu^{exp}-a_\mu^{SM}$ to $2.1$ and $2.3\,\sigma$, respectively.

\appendix
\section{Short-distance constraints}\label{App:SD}
For the parameters contributing to the leading-order chiral LECs \cite{Ecker:1989yg,Pich:2002xy,Weinberg:1967kj,Golterman:1999au,Jamin:2000wn,Jamin:2001zq}:
\begin{equation}\begin{split}
F_V G_V=F^2,&\qquad F_V^2-F_A^2=F^2,\\
F_V^2 M_V^2=F_A^2 M_A^2,&\qquad 4c_d c_m=F^2,\\
8\left(c_m^2-d_m^2\right)=F^2,&\qquad c_m=c_d=\sqrt{2}d_m=F/2.
\end{split}\end{equation}

For the even-intrinsic parity sector \cite{Cirigliano:2006hb,Guevara:2018rhj}:
\begin{equation}\label{SD:eq1}\begin{split}
\lambda_{13}^P&=0,\quad  \lambda_{17}^S=\lambda_{18}^S=0,\\
\lambda_{17}^A&=0, \quad \lambda_{21}^V=\lambda_{22}^V=0.
\end{split}\end{equation}

The analysis of the $\left\langle VAS \right\rangle$ Green function yields \cite{Kampf:2011ty}:
\begin{equation}\label{SD:eq3}\begin{split}
\kappa_2^S=\kappa_{14}^A=0,\quad \kappa_4^V&=2\kappa_{15}^V,\quad \kappa_6^{VA}=\frac{F^2}{32F_AF_V},\\
F_V\left(2\kappa_1^{SV}+\kappa_2^{SV}\right)&=2F_A\kappa_1^{SA}=\frac{F^2}{16\sqrt{2}c_m}.
\end{split}\end{equation}

The study of the $\left\langle VAP\right\rangle$ and $\left\langle SPP\right\rangle$ Green functions yield the following restrictions on the resonance couplings \cite{Cirigliano:2006hb,Cirigliano:2004ue,Cirigliano:2005xn}:
\begin{equation}\label{SD:eq2}\begin{split}
\sqrt{2}\lambda_0 =-4\lambda_1^{VA}-\lambda_2^{VA}-\frac{\lambda_4^{VA}}{2}-\lambda_5^{VA}&=\frac{1}{2\sqrt{2}}\left(\lambda^\prime+\lambda^{\prime\prime}\right),\\
\sqrt{2}\lambda^\prime=\lambda_2^{VA}-\lambda_3^{VA}+\frac{\lambda_4^{VA}}{2}+\lambda_5^{VA}&=\frac{M_A}{2M_V},\\
\sqrt{2}\lambda^{\prime\prime}=\lambda_2^{VA}-\frac{\lambda_4^{VA}}{2}-\lambda_5^{VA}&=\frac{M_A^2-2M_V^2}{2M_VM_A},\\
\lambda_1^{PV}=-4\lambda_2^{PV}=-\frac{F\sqrt{M_A^2-M_V^2}}{4\sqrt{2}d_m M_A},\quad \lambda_1^{PA}&=\frac{F\sqrt{M_A^2-M_V^2}}{16\sqrt{2}d_m M_V}.
\end{split}\end{equation}

For the odd-intrinsic parity sector \cite{Kampf:2011ty}:
\begin{equation}\label{sd:eq1}\begin{split}
\kappa_{14}^V&=\frac{N_C}{256\sqrt{2}\pi^2 F_V},\quad 2\kappa_{12}^V+\kappa_{16}^V=-\frac{N_C}{32\sqrt{2}\pi^2F_V},\quad \kappa_{17}^V=-\frac{N_C}{64\sqrt{2}\pi^2 F_V},\quad \kappa_5^P=0,\\
\kappa^{VV}_2&=\frac{F^2+16\sqrt{2}d_m F_V \kappa_3^{PV}}{32F_V^2}-\frac{N_C M_V^2}{512 \pi^2 F_V^2}, \quad 8\kappa_2^{VV}-\kappa_3^{VV}=\frac{F^2}{8F_V^2}.
\end{split}\end{equation}

\section{Fit results}\label{Fit}
Neglecting all the other contributions, we find
\begin{subequations}\begin{align}
\kappa^{V}_{1}&=(-2.1\pm0.7)\cdot 10^{-2}\text{ GeV}^{-1},\\
\kappa^{V}_{2}&=(-8.8\pm9.1)\cdot 10^{-3}\text{ GeV}^{-1},\\
\kappa^{V}_{3}&=(2.2\pm5.8)\cdot 10^{-3}\text{ GeV}^{-1},\\
\kappa^{V}_{6}&=(-2.1\pm0.3)\cdot 10^{-2}\text{ GeV}^{-1},\\
\kappa^{V}_{7}&=(1.2\pm 0.5)\cdot 10^{-2}\text{ GeV}^{-1},\\
\kappa^{V}_{8}&=(3.1\pm 0.9)\cdot 10^{-2}\text{ GeV}^{-1},\\
\kappa^{V}_{9}&=(-0.1\pm 5.9)\cdot 10^{-3}\text{ GeV}^{-1},\\
\kappa^{V}_{10}&=(-5.9\pm 9.6)\cdot 10^{-3}\text{ GeV}^{-1},\\
\kappa^{V}_{11}&=(-3.0\pm 0.6)\cdot 10^{-2}\text{ GeV}^{-1},\\
\kappa^{V}_{12}&=(1.0\pm 0.8)\cdot 10^{-2}\text{ GeV}^{-1},\\
\kappa^{V}_{13}&=(-5.3\pm 1.1)\cdot 10^{-3}\text{ GeV}^{-1},\\
\kappa^{V}_{18}&=(4.7\pm 0.8)\cdot 10^{-3}\text{ GeV}^{-1}.
\end{align}\end{subequations}
These values are in good agreement with our earlier estimation, $\vert\kappa_i^V\vert \lesssim 0.025\, \mathrm{GeV}^{-1}$ \cite{Miranda:2020wdg}.

\section{Kinematics}\label{App:kin}
For the $\tau^-\to\pi^-\pi^0\gamma\nu_\tau$ decays, we have the following integration region
\begin{equation}\small\begin{split}
\mathcal{D}=&\left\lbrace E_\gamma^{min}\leq E_\gamma\leq E_\gamma^{max},\,x_{min}\leq x\leq x_{max},\,s_{min}\leq s \leq s_{max},-1\leq\cos\theta_-\leq+1,\, 0\leq\phi_-\leq2\pi\right\rbrace,
\end{split}\end{equation}
with boundaries
\begin{equation}\begin{array}{rcl}
\frac{(m_\tau^2-s+x)}{4m_\tau^2}-\frac{\lambda^{1/2}\left(s,x,m_\tau^2\right)}{4m_\tau}\leq& E_\gamma\left(s,x\right)&\leq\frac{(m_\tau^2-s+x)}{4m_\tau}+\frac{\lambda^{1/2}\left(s,x,m_\tau^2\right)}{4m_\tau},\\
4m_\pi^2\leq&s\left(x\right)&\leq\left(m_\tau-\sqrt{x}\right)^2,\\
0\leq&x&\leq\left(m_\tau-2m_\pi\right)^2,\\
\end{array}
\end{equation}
or interchanging the last two limits,
\begin{equation}\begin{array}{rcl}
0\leq& x\left(s\right)&\leq \left(m_\tau-\sqrt{s}\right)^2,\\
4m_\pi^2\leq& s&\leq m_\tau^2.\\
\end{array}\end{equation}
There are other ways to write these,
\begin{equation}\begin{array}{rcl}
4m_\pi^2\leq&s\left(x,E_\gamma\right)&\leq \frac{(m_\tau-2E_\gamma)(2m_\tau E_\gamma -x)}{2E_\gamma}\\
0\leq& x\left(E_\gamma\right)&\leq \frac{2E_\gamma(m_\tau^2-4m_\pi^2-2m_\tau E_\gamma)}{m_\tau-2E_\gamma},\\
E_\gamma^{cut}\leq&E_\gamma&\leq \frac{m_\tau^2-4m_\pi^2}{2m_\tau},\\
\end{array}\end{equation}
or exchanging $x\leftrightarrow E_\gamma$, 
\begin{equation}\begin{array}{rcl}
\frac{(m_\tau^2+x-4m_\pi^2)}{4m_\tau}-\frac{\lambda^{1/2}\left(x,m_\tau^2,4m_\pi^2\right)}{4m_\tau}\leq& E_\gamma\left(s\right)&\leq \frac{(m_\tau^2+x-4m_\pi^2)}{4m_\tau}+\frac{\lambda^{1/2}\left(x,m_\tau^2,4m_\pi^2\right)}{4m_\tau},\\
0\leq& x &\leq (m_\tau-2m_\pi)^2,\\
\end{array}\end{equation}
and
\begin{equation}\begin{array}{rcl}
0\leq&x\left(s,E_\gamma\right)&\leq \frac{2E_\gamma(m_\tau^2-s-2E_\gamma m_\tau)}{m_\tau-2E_\gamma}\\
4m_\pi^2\leq& s\left(E_\gamma\right)&\leq m_\tau(m_\tau-2E_\gamma),\\
E_\gamma^{cut}\leq&E_\gamma&\leq \frac{m_\tau^2-4m_\pi^2}{2m_\tau}.\\
\end{array}\end{equation}
Further, interchanging $s\leftrightarrow E_\gamma$, we get
\begin{equation}\label{Appx4:eq70}\begin{array}{rcl}
E_\gamma^{cut}\leq&  E_\gamma\left(s\right)&\leq \frac{m_\tau^2-s}{2m_\tau},\\
4m_\pi^2\leq& s &\leq m_\tau(m_\tau-2E_\gamma^{cut}).\\
\end{array}\end{equation}

\bibliographystyle{JHEP}
\bibliography{bibl}

\end{document}